\documentclass{elsart}

\usepackage{natbib}

\usepackage{graphicx}

\usepackage{amssymb}
\usepackage{amsmath}
\usepackage{amsfonts}

\begin{document}

\lefthyphenmin=2
\righthyphenmin=2

\renewcommand{\topfraction}{1}
\renewcommand{\bottomfraction}{1}

\begin{frontmatter}

\title{Regulation of Neuromodulator Receptor Efficacy - 
Implications for Whole-Neuron and Synaptic Plasticity}

\author{Gabriele Scheler}

\address{International Computer Science Institute\\ 
1947 Center Street, Suite 600\\ Berkeley, Ca. 94704}

\ead{scheler@icsi.berkeley.edu}


\date{}

\begin{abstract}
Membrane receptors for neuromodulators (NM) are highly regulated in their 
distribution and efficacy - a phenomenon which influences the individual
cell's response to central signals of NM release. 
Even though NM receptor regulation is implicated in the pharmacological 
action of many drugs, and is also known to be influenced by various
environmental factors, its functional consequences and modes of action 
are not well understood.
In this paper we summarize relevant experimental evidence on NM receptor 
regulation (specifically dopamine D1 and D2 receptors) in order to 
explore its significance for neural and synaptic plasticity.
We identify the relevant
components of NM receptor regulation (receptor phosphorylation, receptor 
trafficking and sensitization of second-messenger pathways) 
gained from studies on cultured cells.
Key principles in the regulation and control of short-term plasticity 
(sensitization) are identified, and a model is presented which employs
direct and indirect feedback regulation of 
receptor efficacy. We also discuss long-term plasticity which involves 
shifts in receptor sensitivity and loss of responsivity to NM signals.
Finally, we discuss the implications of NM receptor regulation for models 
of brain plasticity and memorization.
We emphasize that a realistic model of brain plasticity will have to go
beyond Hebbian models of long-term potentiation and depression.
Plasticity in the distribution and efficacy of NM receptors may provide 
another important source of functional plasticity with implications
for learning and memory.
\end{abstract}


\end{frontmatter}

\section{Introduction}

G-protein coupled receptors (GPCRs) which comprise a major group of 
cellular receptors 
on many types of cells, including neurons, undergo significant plasticity. 

GPCRs become desensitized (phosphorylated) as a form of short-term 
plasticity, which means that receptors 
are temporarily uncoupled from their effectors (G-proteins). They also 
become down- or upregulated in a more lasting form of plasticity, which 
involves receptor trafficking between intracellular stores and the cell 
membrane, and in some cases receptor degradation as well as 
new protein synthesis.
These processes affect receptors for the main neuromodulators serotonin,
dopamine, noradrenaline and acetylcholine, for neuropeptides such as 
opioids, 
for neurohormones such as steroids or estrogen, 
as well as the metabotropic glutamate (mGLU) and GABA$_B$ receptors in 
the brain.
 
In general, receptors are sensitized and desensitized in response to agonist 
exposure, modulated by cell internal parameters and synaptic activation.
Important parameters for receptor plasticity are cell-internal calcium,
and the second-messenger dependent protein kinases A and C (PKA and PKC), as 
well as G-protein specific kinases (GRKs).
The time-scale of these changes is within minutes for desensitization
and several hours for alterations in receptor distribution,
which is comparable to 'early' and 'late' long-term potentiation.

The functional significance of this adaptive 
regulation is at present not well understood.
Receptor regulation has mainly be studied in response to various 
pharmacological agents (antipsychotics, antidepressants, drugs of abuse),
where sensitization of NM responses has consistently been shown in 
different tissues and cell types (e.g. ventral tegemental area and 
nucleus accumbens)
\cite{HenryWhite95,WangWetal96}.
Behavioral effects of stress \cite{ScheggiSetal2002}, learning 
(inhibitory avoidance, 
\cite{OrtegaAetal96,vanderZeeLuiten99,vanderZeeEAetal94}), 
and environment (novelty, social conditions) have also been documented
\cite{MorganDetal2002,LanfumeyLetal99}.

This evidence which points at an experience-dependent regulation of NM 
receptors coexists with a significant body of data showing 
constitutive expression of receptors to different types of neurons.
The level of receptor expression varies for projection neurons 
(cortical pyramidal cells or striatal medium spiny projection neurons) vs. 
local interneurons (fast-spiking vs. regular spiking neurons) 
\cite{JakabGoldman-Rakic2000,LeMoineGaspar98},
for different cortical layers or for patches in amygdala or striatum
\cite{vanderZeeLuiten99},
according to neuropeptide colocalization (e.g., substance P, enkephalin)
\cite{SurmeierDJetal96},
and for neurons with a different connectivity (striato-nigral vs. 
striato-pallidal neurons) \cite{YungBolam2000}. 
There is also developmental regulation of receptor expression 
which is different postnatally \cite{MengSZetal99,GurevichEVetal2001}, during 
adolescence \cite{MontagueDMetal99,TeicherMHetal95}, as well as during 
ageing \cite{PowerJMetal2002}.

Constitutive receptor distribution sets 
the boundaries within which ex\-perien\-ce-dependent fluctuations occur. 
These fluctuations may be 
{\it transient} or they may have a {\it permanent} component corresponding 
to short-term desensitization and long-term down- or up-regulation.
In this paper we provide an overview of the biological mechanisms involved 
in NM receptor regulation with the goal of analyzing the adaptive function 
of this process.
We will find that receptor efficacy undergoes significant changes that 
are important in mediating neuromodulatory signals. We will also find 
that the regulatory processes underlying receptor plasticity are partly 
overlapping with the processes underlying LTP/LTD.  
 
We suggest that NM receptor regulation is a process which has the capacity 
to contribute to brain plasticity on the population level, on the level of 
the individual neuron and probably also on the level of the synapse.
This provides a mechanism for memorization and an added storage capacity 
beyond Hebbian long-term potentiation and long-term depression.
A better insight into the role of NM receptor regulation may 
lead to a new understanding of brain 
plasticity and a thorough revision of current theories of memory and learning.

\section{Protein Regulatory Networks Underlying Receptor Plasticity}

\subsection{Component processes of receptor regulation}
The molecular biology of receptor regulation 
has been elucidated in considerable detail, mostly
by studies on cultured cells stably transfected with receptors 
at fixed concentrations.

A number of different components have been identified:
\begin{enumerate}
\item conformational change of receptor protein and functional uncoupling 
from effectors (G-proteins) by {\bf phosphorylation}.
\item translocation of receptors from the membrane to a cytoplasmic structure
(in endoplasmic reticulum and Golgi apparatus) ({\bf receptor internalization}).
\item reduction in potency and efficacy of a receptor in inducing 
changes in second-messenger concentration (adenylyl cyclase, cAMP)
({\bf desensitization}).
\item {\bf receptor degradation} in lysosomes, which removes 
receptor proteins from the cytoplasma as well as the plasma membrane.
\item translocation from a pool of internally stored receptors to the 
membrane ({\bf recruitment}, resensitization).
\item delivery of newly synthesized receptors 
to the membrane ({\bf protein synthesis}).
\end{enumerate}

This shows that receptor regulation is far from being a simple process, as 
might be expected if homeostatic regulation by feedback control were its
only function. Rather, the "layering" of several processes indicates that 
both short-term and long-term regulation of receptors occurs and that  
different "entry-points" for interacting processes exist to influence the 
outcome of a specific stage in receptor regulation.

The prototype case for GPCR regulation has been the regulation 
of the $\beta$-adrenergic receptor 
\cite{KrupnickBenovic98,HausdorffWPetal90},
but the processes involved are somewhat different for each individual 
GPCR type. 
We focus here on the dopamine D1 receptor, which is also well documented, 
and which is of considerable significance in the regulation of 
membrane excitability and neural signal transmission.
 
\subsection{Protein regulatory network}

\begin{figure}[htb]
\begin{center}
\ \includegraphics[width=0.9\textwidth]{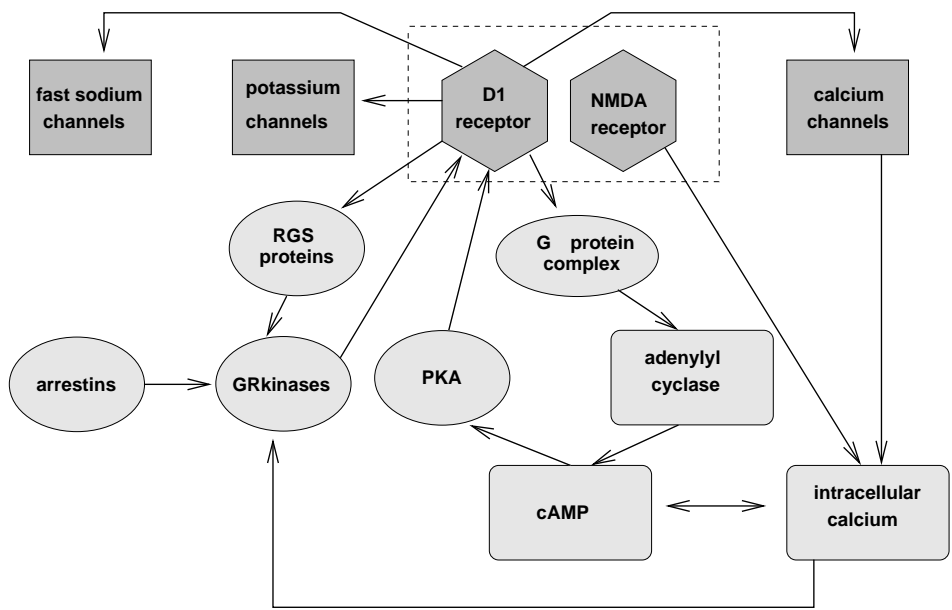}
\caption{D1 receptor effectors and regulators: dark shaded, membrane receptors 
and ion channels, light shaded, intracellular pathways. Possible dimerization 
D1-NMDA is shown. Note the direct feedback regulation via PKA, also via RGS,
GRK and the indirect regulation via calcium, GRK. Possible interactive effects 
between cAMP and Ca are also shown.}
\label{variables2}
\end{center}
\end{figure}

The specific proteins and signalling substances involved in receptor 
regulation are shown as a regulatory network for the dopamine D1 
receptor in Fig.~\ref{variables2}. 

Functional D1 receptors are coupled to G-proteins, which form heterotrimeric 
complexes.
When receptors become activated, the coupled G-proteins dissolve into two 
components: G$_{\alpha}$ and G$_{\beta\gamma}$.
These components have the ability to modulate a number of membrane 
ion channels, such as inwardly rectifying K+ channels (e.g. GIRK channels)
\cite{MomiyamaKoga2001},
$I_{h}$ and other K+ channels \cite{NicolaSMetal2000},
high-voltage gated calcium channels (L-type, P/Q/R-type and N-type calcium 
channels) \cite{Hernandez-LopezSetal97,SurmeierDJetal95},
and also sodium channels \cite{CantrellARetal99,MauriceNetal2001}.
D1 receptors furthermore have the effect of increasing NMDA transmission
\cite{WangODonnell2001,Flores-HernandezJ1etal2002}, and
may have complex effects on local calcium levels \cite{KassackMUetal2002}.

D1 receptors communicate with cell-internal pathways by activating 
$G_{olf}$-proteins \cite{ZhuangXetal2000}, which raise adenylyl cyclase and cAMP-levels.
cAMP-levels and calcium levels are required for the activation of the 
transcription factor CREB and the 'early genes' fos and delta-fosB.
Both proteins regulate 
the translational activity of mRNA and are critically involved in 
new protein synthesis.
D1 receptor coupling is reduced by protein kinases - both 
cAMP-dependent kinase PKA and G-protein specific kinases (GRKs)
\cite{KrupnickBenovic98,FergusonSS2001}.
These kinases phosphorylate the receptor protein and contribute to the 
desensitization of its effect on second messengers.
Calcium indirectly reduces desensitization, via the calcium-dependent 
protein calmodulin, which inhibits GRKs \cite{CarmanBenovic98}.
Another pathway for calcium to prevent desensitization has recently been 
detected in the calcium sensor NCS-1, which reduces phosphorylation and 
internalization of the D2 receptor \cite{KabbaniNetal2002}.
Arrestins can increase the effects of GRKs, for 
instance, overexpression of arrestins reduces the ability of 
$\beta$-adrenergic receptors to activate $G_S$ by $>75\%$ 
\cite{LohseMJetal90}. They play a major role
in the internalization of receptors. 

Desensitization is furthermore influenced by RGS-proteins, which
regulate G-proteins and G-protein signaling by 
activating GTPase \cite{JeongIkeda2000,vonZastrowMostov2001,ZhengBetal2001}.
GTPase is the kinase which phosphorylates G-proteins, and renders them 
insensitive to receptor activation.
RGS proteins take part in producing fast kinetics in vivo
by favoring reformation of the heterotrimeric 
state (G$_{\alpha}$ + G$_{\beta\gamma}$),
while the hydrolysis of GTPase is 40-fold slower in the absence of RGS 
\cite{JeongIkeda2000}.
The effect of overexpression of RGS proteins is a change in kinetic rate, 
an acceleration of the desensitization-resensitization cycle and also 
a net decrease of desensitization.
RGS levels themselves may become upregulated by dopamine D1 and D2 activation 
e.g. in striatum, specifically  RGS-2 and RGS-4 seem to be enhanced by 
D1 or D2 receptor activation respectively 
\cite{GeurtsMetal2002,TaymansJMetal2003}.

\section{Short-Term Desensitization of Receptors}
\subsection{Receptor phosphorylation and desensitization of second-messenger 
activation}
\label{3.1}
Receptor efficacy in general is determined by the amount of functional 
coupling of an agonist and the reactivity of effector pathways.

Receptor phosphorylation and internalization affect signal transduction by 
agonist binding as a form of short-term variation.
Membrane receptors undergo functional decoupling by phosphorylation at multiple
Ser and Thr residues, which induces conformational change of the protein and 
prevents effective ligand binding \cite{JiangSibley99,MasonJNetal2002}.
Phosphorylation is fast, with a half-time of less than a minute for the 
D1 receptor and it is also reversible upon agonist removal with a half-time 
for resensitization of about 10-15 minutes for the D1-receptor
\cite{GardnerBetal2001,VickeryvonZastrow99}, cf. \cite{MorrisonKJetal96}.
%

Phosphorylation and internalization may be achieved by 
PKA, which 
is cAMP-dependent, or by GRKs, which are subject to regulation by calcium 
via  calmodulin or NCS-1 \cite{ChuangTTetal96,HagaKetal97,ProninANetal97}.
Calcium may enhance receptor efficacy, when
intracellular calcium binds to calmodulin, and inhibits 
GRKs, e.g. GRK5 {\cite{ProninANetal97}.
Interestingly, there may be another, "corrective", pathway for a direct 
interaction between calcium and GRK: 
the inhibition of GRKs by 
calmodulin can be 
abolished by high levels of (calcium-dependent) protein kinase C (PKC)
\cite{KraselCetal2001}. 
In general, overexpression 
of GRKs leads to reduced functional coupling and requires more agonist 
to achieve the same amount of effect on second messengers (subsensitivity)
\cite{BouvierMetal88,NgGYetal94,TiberiMetal96}.

The role of PKA in phosphorylation of the D1 receptor is 
to a certain degree controversial \cite{SibleyDRetal98}. 
There is indirect evidence from cell lines with reduced PKA activity 
where desensitization is attenuated
\cite{VenturaSibley2000}, and from mutant receptors which lack a 
PKA phosphorylation site, where 
the onset time of desensitization is greatly reduced \cite{JiangSibley99}.
Also, \cite{BatesMDetal91} and \cite{BlackLEetal94} show that stimulating PKA 
can mimic agonist-dependent desensitization.
But there
are also data by \cite{LewisMMetal98,BatesMDetal93}, which seem to indicate 
that PKA is not important in desensitization.
The work by \cite{MasonJNetal2002} 
(based on a PKA-insensitive mutant receptor)
suggests that PKA has a major effect on 
receptor trafficking within the cell and increases proteolysis, but 
is not strictly required for desensitization of cAMP.
Thus it has 
been suggested that PKA modifies a later process in desensitization, which 
is more intimately linked to internalization and recycling probability, rather 
than agonist-induced phosphorylation. In this sense, GRKs and PKA are 
most effective at different stages of the desensitization process. This 
would also imply a different time course of their feedback regulation, since 
PKA would operate with a longer delay in its reduction of receptor efficacy.

%

%


\subsection{Key factors in desensitization}

When a receptor is in a phosphorylated state, it is effectively uncoupled 
from its effectors, until it becomes dephosphorylated. For the dopamine 
D1 receptor, both of these processes can be performed without internalizing
the receptor.
However internalization is often the consequence of phosphorylation and both 
processes together may be termed "desensitization", since they affect the 
functional efficacy of a receptor population in transmitting signals to 
intracellular pathways.

\begin{figure}[htb]
\noindent
\begin{center}
\includegraphics[width=11cm]{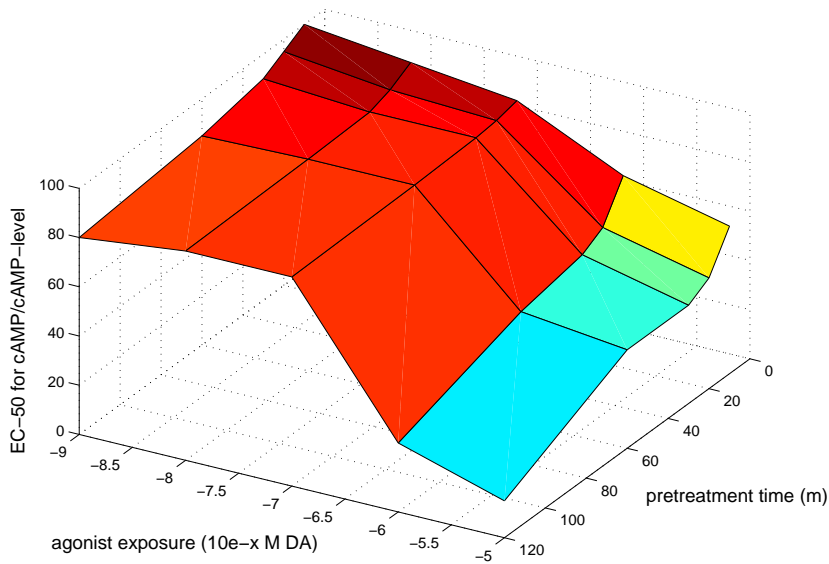}
\end{center}
\caption{Desensitization of the 
cAMP pathway for the D1 receptor. Data have been  
combined from 
\cite{NgGYetal95}, \cite{GardnerBetal2001} and \cite{LewisMMetal98}. 
The sigmoid shape of the dose-response relationship is clearly evident 
in the vertical axis. Length of pretreatment in contrast seems to follow 
a logarithmic curve. The reduction of efficacy can reach up to 80\%.}
\label{d1real}
\end{figure}

To study the desensitization mechanism for a given receptor, receptors are 
overexpressed in cultured cells, and then exposed to agonists at different 
concentrations and for different times. The amount of functional coupling 
is assessed by measuring concentrations of adenylyl cyclase or cAMP.
These experiments have shown that functional efficacy strongly 
depends on the level of agonist exposure. 

Fig.~\ref{d1real}, \ref{d2real} show the time course of desensitization by measuring 
the concentration of adenylyl cyclase and cAMP. 
In Fig.~\ref{d1real} data from 
\cite{NgGYetal95}, \cite{GardnerBetal2001} and \cite{LewisMMetal98} are 
combined.
The desensitization of the D1 receptor reduces the ability of a brief dopamine 
challenge to enhance adenylyl cyclase and cAMP-levels. This desensitization 
is both dose-dependent and dependent on pretreatment time. Dose-dependence follows
a sigmoidal shape, with a critical range between $10^{-6}$ and $10^{-7}$ mol dopamine.
Pretreatment time does not increase desensitization much beyond an initial effect.
Overall, there is a reduction in receptor efficacy of up to 80\%.

\begin{figure}[htb]
\noindent
\begin{center}
\includegraphics[width=11cm]{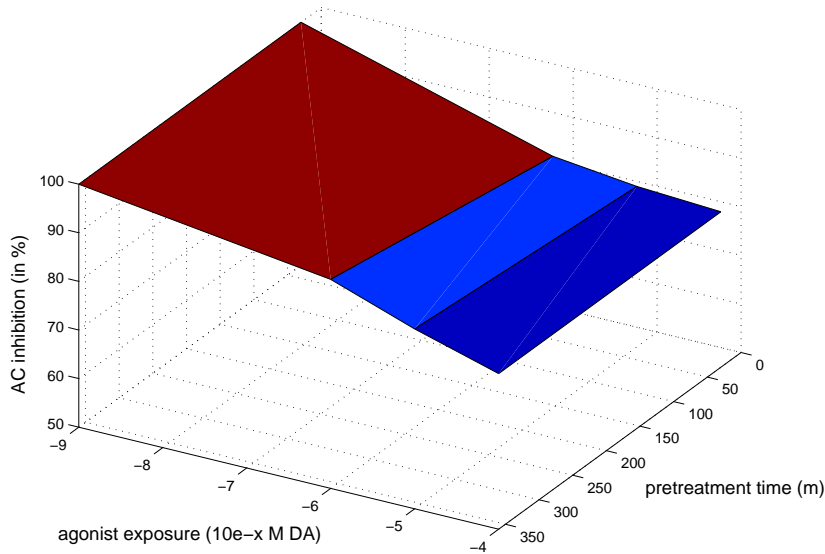}
\end{center}
\caption{Desensitization of adenylyl 
cyclase(AC) inhibition for the D2 receptor. Data are taken 
from \cite{NgGYetal97}. Possibly sigmoidal and logarithmic shapes 
of the curves are less evident, probably as a result of the limited data 
set. The effect of desensitization is limited to about 20\%.}
\label{d2real}
\end{figure}

Fig.~\ref{d2real} shows similar data from the desensitization of 
the D2 receptor \cite{NgGYetal97}. 
The D2 receptor has the capacity to inhibit the adenylyl cyclase/cAMP pathway 
and this capacity is reduced by agonist-dependent desensitization only up
to 20\%.  The critical range of agonist exposure, however, is similar. Again, 
pretreatment time (here up to 6 h) is not a strong factor.

We can derive a general functional form for receptor efficacy (C)
which reflects the dependence on agonist stimulation (A):
\begin{equation}
 C = {1 \over 1+e^{-A}} . 
\end{equation}

The amount of agonist stimulation can be described by an integral over 
the NM concentration:
\begin{equation}
 A = \int NM_t dt . 
\label{agonist}
\end{equation}

The sigmoidal shape of the function reflects the fact that receptor efficacy
is almost linearly dependent on agonist exposure within a certain 
concentration range 
and reaches saturation or stays below a threshold otherwise.
This is visualized in Fig.~\ref{desens}. Furthermore, Fig.~\ref{desens} shows 
the modification of this basic shape by additional parameters (s. below).

\begin{figure}[htb]
\begin{center}
\ \includegraphics[width=11cm]{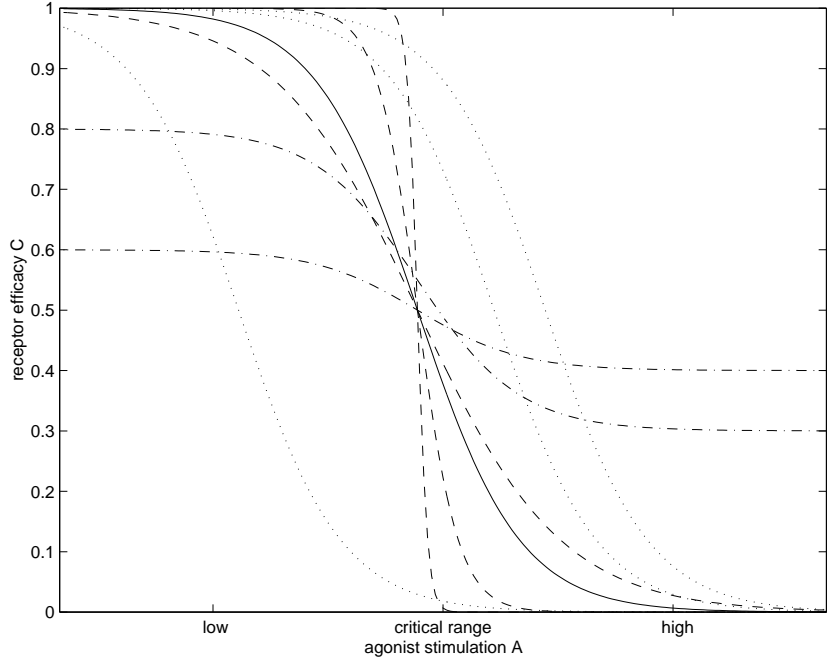}
\caption{General form of NM  receptor efficacy and its variability by 
parametrization. Dotted lines (\ldots) show shifts in dose-dependence ($I_1$),
dashed lines (-\ -) correspond to the rate of desensitization ($I_2$), 
dashed-dotted lines (-.) correspond to the degree of agonist-dependence
($\lambda$).}
\label{desens}
\end{center}
\end{figure}

It may well be that receptor desensitization depends on the particular 
time course of agonist exposure rather than the total amount. 
These experiments, though technically feasible, have not systematically 
been carried out.
For instance,
phasic increases of NM concentrations which are short-lasting may fail to 
desensitize receptors significantly, while tonic increases with a smaller 
total amount of agonist stimulation may have a larger effect. However, 
in the absence of experimental data, integrating over agonist concentrations 
in time seems to be the best approximation. Further work may allow to refine 
the formula in eq.~\ref{agonist} to incorporate different temporal patterns 
of agonist stimulation.

This basic agonist-dependent desensitization may be modified by a number 
of factors.
First of all, as we have seen, there are several cell-internal parameters
which influence the magnitude of desensitization. 
For instance, a high level 
of GRKs  shifts the agonist-dependence curve to the right
\cite{BouvierMetal88,NgGYetal94,TiberiMetal96}.
A basic parameter 
$I_1$ allows to express sub- and supersensitivity of receptors, defined 
by dose-dependence of agonist exposure:

\begin{equation}
 C = {1 \over 1+e^{-A+I_1}} . 
\end{equation}
Presumably, 
the amount of phosphorylation by protein kinases can be expressed with 
this parameter. 
Experimental data show that $I_1$ can be  
manipulated by overexpression of GRK \cite{TiberiMetal96}.
We'd similarly 
expect increased calcium/calmodulin to have an effect on this parameter.


The action of RGS proteins is different in that it has a significant effect 
on the resensitization dynamics. 
Receptors prompt conversion of the inactive G protein to an active form.
RGS proteins accelerate the conversion of the activated G protein
back to its inactivated 
form \cite{vonZastrowMostov2001,ZhengBetal2001}.
Several studies have shown \cite{ChenLambert2000,JeongIkeda2000} that 
reduced RGS activity leads to significantly less receptor efficacy both in 
terms of slow onset and prolonged recovery times, measured by N-type calcium 
channel or GIRK channel activity (see Fig.~\ref{rgs}).
In contrast, high RGS levels lead to fast recovery and may reduce 
onset of receptor activation \cite{JeongIkeda98,SaitohOetal97,DoupnikCAetal97,
WatsonNetal96}.

\begin{figure}[htb]
\begin{center}
\ \includegraphics[width=11cm]{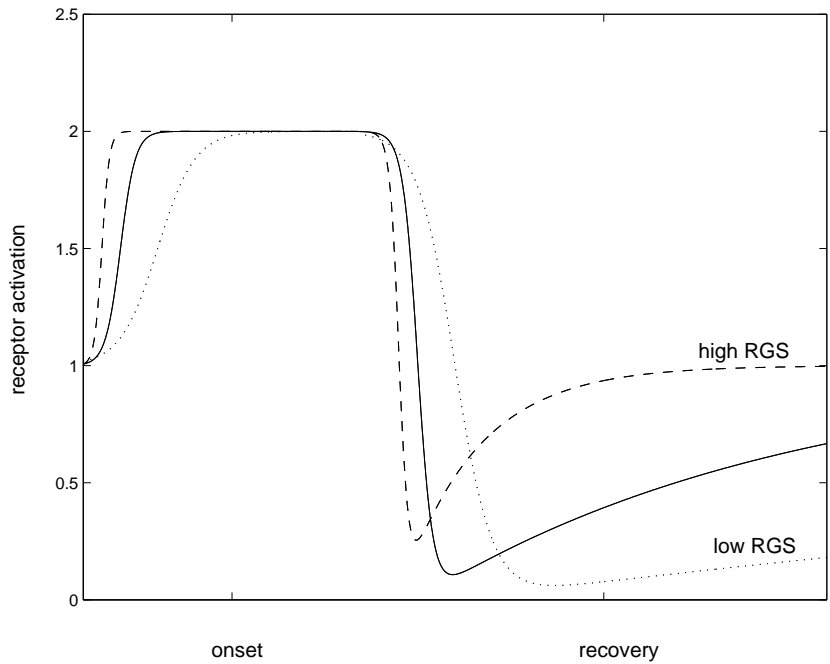}
\caption{Desensitization dynamics modulated by RGS-proteins. The functional
shape is adapted from \cite{JeongIkeda2000} and \cite{JeongIkeda98}, 
according to alterations in the half recovery time $t_{0.5}$ and the rise 
time $t_{rise}$. }
\label{rgs}
\end{center}
\end{figure}

Thus RGS proteins speed up resensitization and affect receptor efficacy 
independent of agonist levels.  
This effect can be expressed by a parameter $I_2$:

\begin{equation} \label{eq-final}
 C = {1 \over 1+e^{{-A +I_1} \over I_2}} . 
\end{equation}

Finally, desensitization depends on the amount of time a receptor 
exists in the functional conformational state. 
Receptors may exhibit a high degree of spontaneous conformational 
change, which means that they become activated even in the absence of agonist 
binding.
For dopamine receptors, D5 receptors, which are similar to D1 receptors, have 
a high degree of this constitutive activity \cite{TiberiCaron94}, cf. 
\cite{GrotewielSanders-Bush99,DemchyshynLLetal2000}.
This can be expressed by a factor for desensitization ($\lambda$) which is 
constitutive for each receptor type.

\begin{equation}
 C = \lambda {1 \over 1+e^{{-A +I_1} \over I_2}} . 
\label{efficacy-eq}
\end{equation}

Receptor efficacy as measured by effects on cAMP/adenylyl cyclase need not
be identical to effects on membrane ion channels and consequently neural 
transmission.
But in general, dopamine D1 mediated effects on membrane excitability can be 
mimicked by forskolin-mediated stimulation of PKA, implicating the 
cAMP-pathway as the major factor in regulating membrane properties. 
The cAMP/adenylyl cyclase pathway thus provides an important 
state parameter for the regulation of protein phosphorylation, the expression 
level of protein kinases and a common pathway for a number of G-protein 
coupled receptors. 
Strictly synaptic components of receptor 
activation (NMDA channel regulation and presynaptic regulation of transmitter 
release), however, may be regulated by other pathways, with prominent 
dependence on calcium and protein kinase C levels.
The cAMP-pathway also influences 
early gene expression (c-fos, delta-fosB, CREB),
which are important variables for protein synthesis and any functional
cell plasticity that is regulated in the long-term.
Thus we can regard regulation of the cAMP pathway as a basic mechanism that
may affect receptor efficacy for both membrane excitability and gene expression/
protein synthesis (cf. \cite{LameyMetal2002,GardnerBetal2001}).
The relation between receptor efficacy and synaptic modulation  
may be regulated in a somewhat different way.

We have seen that the efficacy of the D1 receptor in raising intracellular 
cAMP-levels depends on the influence of different 
factors on the desensitization function.

Three different types of parameter 
($\lambda$, $I_1$, $I_2$) can be 
distinguished by their influence on receptor efficacy (see Fig.~\ref{desens}).
\begin{itemize}
\item[]
$I_1$ shifts the function to the left or right without affecting its 
shape, corresponding to an alteration of the dose-response relationship. 
\item[]
$I_2$ alters the steepness of the function and thus the temporal properties 
of receptor desensitization.
\item[]
$\lambda$ flattens the curve indicating less dependence on agonist stimulation.
\end{itemize}

This basic variability of receptor efficacy emerges from the 
parametrization of the functional form. 
Parameter fitting to experimental data may allow a further quantitative 
analysis of these functional relations.

\subsection{Feedback loops in regulating receptor efficacy}

In this section, we will outline a model for the dynamic interactions that
determine receptor efficacy.
We will develop a qualitative approach that demonstrates the emergence 
of bistability in receptor efficacy - understood here as a stable state 
which results from a prolongation of the cellular response beyond the actual 
duration of the signal. 
In a slightly different sense, bistability emerges when there are trigger 
signals both to induce a new state and to turn it off
(cf. \cite{BhallaIyengar2001,Bhalla2002,WengGetal99}
for discussions of bistability and multistability in this context),
and as we shall see this condition is probably also met.
The model is purely qualitative, an attempt to illustrate the conditions 
that might underlie a short-term regulation of receptor efficacy that 
provides an important physiological state parameter for understanding 
the computational properties of neuromodulation. But a full numerical 
model, e.g. based on chemical mass-action models \cite{MishraBhalla2002,
LeeEetal2003} could be developed on the basis of this approach and 
would allow more detailed investigations of these interactions.

\begin{table}
\begin{center}
\begin{tabular}{|lll|}
\hline
&&\\
\ \ &$V_{PKA}= \alpha_1 \ I_{PKA} + \beta_1 \ V_{NM}$&\\
&$V_{RGS}= \alpha_4 \ I_{RGS} + \beta_4 \ V_{NM} $&\\
&$V_{Ca} = \alpha_3 I_{Ca}+ \beta_3 \ V_{NM}$&\\
&$V_{GRK} = \alpha_2 I_{GRK}- \beta_2 \ V_{Ca}$&\\
&$I_1 = -\gamma_1 \ V_{PKA} - \gamma_2 \  V_{GRK} + \gamma_3 V_{RGS}$&\ \ \\
&$I_2 = \gamma_4 V_{RGS} $&\\
&$A = \sum_{t=0} I_{NM}$&\\ 
&$C = 1 / (1+e^{(-A +I_1)\ 1/I_2})$& \\
&$V_{NM} = C \ I_{NM}$&\\
&&\\
\hline
\end{tabular} 
\end{center}
\caption{A system of equations defines a simulation model for 
the regulation of receptor efficacy $C$. Concentrations for PKA, Ca 
and RGS ($V_x$) are determined by an independent component $I_x$ and 
by feedback from the receptor activation $V_{NM}$. GRK is determined by 
$I_{GRK}$ and the calcium level $V_{Ca}$. $\lambda$ is regarded as a fixed 
parameter for the $D_1$ receptor. The actual receptor activation $V_{NM}$ 
is modeled as a product of agonist concentration and receptor efficacy.}
\label{table1}
\end{table}

The model determines the parameters $I_1$ and $I_2$ in eq. \ref{efficacy-eq} 
from kinase activity ($I_1$) and RGS-level ($I_2$). 
The protein regulatory interactions that determine the parameters $I_1$ and 
$I_2$ in this model are  described by a number of equations 
as shown in Table \ref{table1}. 
Each of the concentrations in the model is calculated from a component that is 
independent of feedback within the system 
($I_{Ca}$, $I_{PKA}$, $I_{GRK}$ and $I_{RGS}$), and 
a component that depends on the activation of the receptor ($V_{NM}$).
GRK levels specifically are influenced by indirect feedback via calcium
levels - which provides an entry-point for a more complex regulation 
of GRK levels by other processes affecting calcium.
The ratio between these components is determined by the factors $\alpha_i$
and $\beta_i$.
This analytical approach of separating system-internal components from outside 
influences has clear advantages in model construction and control over 
system components and also enables modular construction with clearly 
defined interfaces which is lacking in more 'bottom-up' approaches to 
intracellular modeling \cite{LeeEetal2003,WileyHSetal2003}.

However, it is assumed that the effect of the receptor activation $V_{NM}$ on the 
concentration is only linear - raising or lowering levels by an amount 
proportional to $V_{NM}$. This is a simplifying assumption that 
may have to be examined in further work.

\begin{figure}[htb]
\noindent
%
\centerline{\includegraphics[width=0.8\textwidth]{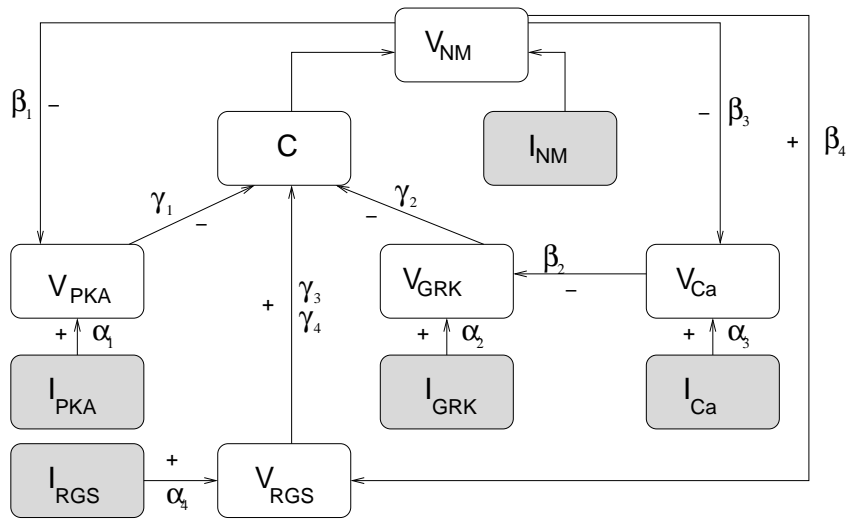}}
\caption{The diagram shows the dependencies between the concentrations 
for the kinases GRK and PKA, for calcium, and the role of RGS proteins.
The factors $\alpha_i$, $\beta_i$ and $\gamma_i$ are shown as labels 
on the links between parameters.
}
\label{linear-feedback}
\end{figure}

Fig.~\ref{linear-feedback} visualizes the relations between the parameters.
The target regulated value is the receptor efficacy $C$. The actual 
receptor activation effect $V_{NM}$ is defined by both the efficacy C and 
the current agonist stimulation $I_{NM}$.
The main parameters $I_1$ and $I_2$ are defined by 
concentrations of the kinases PKA and GRK ($V_{PKA}$ and $V_{GRK}$), and 
the levels of RGS-proteins ($V_{RGS}$).
Specifically, $I_1$, which determines the right- or left shift of 
the dose-response curve for the receptor,  is assumed to be a linear 
combination of the concentrations for the  two kinases. Since these 
kinases may act with different time courses, a more complex interaction 
may have to be empirically determined by more detailed experimental work. 

$C$ is being set by the agonist exposure summed over time, and indirectly 
by the PKA, GRK and RGS levels. 
There is also some experimental evidence concerning the free parameters 
$\alpha_i$, $\beta_i$ and $\gamma_i$ in the system. For instance,
the ratio $\beta_1$/$\alpha_1$ should be larger 
than $\beta_3$/$\alpha_3$. 
This is the case, since the contribution of D1 receptor activation to PKA 
levels is probably much higher than its contribution to calcium levels 
\cite{KassackMUetal2002,NgGYetal95,GardnerBetal2001,LewisMMetal98}. 
Also,  $\gamma_2 > \gamma_1$, since GRK seems to have a stronger effect 
than PKA on receptor regulation (see above, section~\ref{3.1}).

With this system we arrive at a simplified, but 
functional model for the computation of receptor
activation, taking the internal state of the neuron, defined by protein and 
calcium concentrations, into account. 
Further work is needed to more fully develop such a model as a basis for 
empirical predictions and in close correspondence with ongoing experimental
work (cf. \cite{LeeEetal2003} for an example of such experimental validation).

However, even the basic model can be employed to illustrate the 
emergence of a limited form of bistability, which may provide an important 
prerequisite for functional long-term plasticity as memory for brief signals 
to appear.

Receptor efficacy undergoes both negative (cAMP-dependent) 
and positive (calcium-dependent, RGS-mediated) feedback.
There are two main regulatory loops, the PKA/PKC-mediated and 
GRK-mediated feedback. 
PKA/PKC are up-regulated by D1, and thus 
provides a {\it negative} feedback loop.
GRK levels are influenced by internal calcium and thus constitute a 
{\it positive} feedback loop.

In general negative feedback leads to oscillatory dynamics with appropriate 
phase delays, while additional positive feedback can lead to the 
establishment of multistability \cite{Bhalla2003}.
 
The presence of antagonistic feedback effects suggests bistability mediated by 
calcium in the presence of slow oscillatory rhythms regulated by PKA.
A simulation of the system shows the emergence of these features.
The graph in Fig.~\ref{simul1}
shows the induction and persistence of a state of high 
receptor efficacy by two calcium spikes.

\begin{figure}[htb]
\noindent
\begin{center}
\ \includegraphics[height=6cm,width=7.5cm]{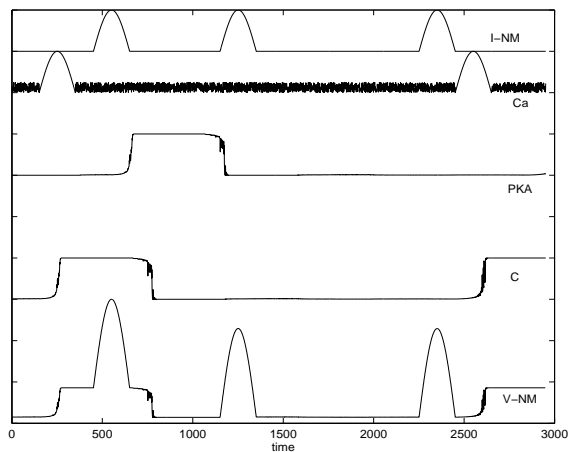}
\end{center}
\caption{Calcium-induced multistability of receptor efficacy. This simulation 
uses the model defined in Table~\ref{table1}, with a time delay for 
$V_{PKA}$ ($\alpha_1 I_{PKA} + (\beta_1 V_{NM} -delay))$).
Brief triggering increases in calcium concentrations $I_{Ca}$ induce changes 
in receptor efficacy $C$ maintained by positive feedback and switched off 
by delayed increase in PKA-level. Actual receptor activation V-NM has different
magnitudes for identical signals $I_{NM}$.}
\label{simul1}
\end{figure}

In the first case, the increase in calcium precedes a neuromodulatory
signal. Due to feedback effects, the increase in receptor efficacy $C$ 
persists even when the calcium signal is terminated. Accordingly, the response 
of the receptor to this signal is increased compared to an agonist signal
in the absence of calcium (second signal). This model also shows a critical 
dependence of response on the respective timing of Ca and $I_{NM}$ signals, 
which should be testable empirically.

The existence of bistable or multistable states that persist beyond the 
duration of a trigger signal shows the realization of an important 
computational principle, the equivalent of a flip-flop device.
This is a basic memory element and probably a necessary component 
for long-term plasticity of NM receptor efficacy. The critical dependence 
of efficacy on relative timing of calcium and NM signals opens up 
interesting perspectives for an interaction of NMDA and D1 related signals.

\section{Long-term Plasticity of Neuromodulatory Receptors}
\subsection{Receptor super- and subsensitivity}
\label{sensitivity}
Besides the short-term processes of densensitization and resensitization, 
there are also distinct processes of long-term regulation of NM receptor 
efficacy \cite{TsaovonZastrow2000}.
Long-term alterations in receptor efficacy for membrane-bound responses can be 
measured directly by responsivity to agonist stimulation.

Both subsensitive and supersensitive populations of cells have been 
described in various tissues.
Dopamine D1 supersensitivity has been shown in striatum after interruption 
of the dopaminergic nigro-striatal pathway 
\cite{CaiGetal2002} or in genetically altered dopamine-deficient mice 
\cite{KimDSetal2000}.  
Exposure and withdrawal conditions for amphetamine or cocaine also change
dopamine receptor sensitivity.
For instance, there is D1 receptor supersensitivity in 
nucleus accumbens 
\cite{UnterwaldEMetal94,HenryDJetal98,MayfieldRDetal92,HenryWhite95,HuXTetal2002}
and D2 receptor subsensitivity
in nucleus accumbens \cite{ChenJCetal99,HenryDJetal98}
and ventral tegmental area \cite{HenryDJetal98}.
Treatment with antipsychotic medication has also shown consistent shifts 
in receptor sensitivity in a number of brain areas (ventral tegmental area,
prefrontal cortex, basal ganglia)\cite{SeeREetal90,TaraziFIetal97,LeeHetal99}. 
Furthermore, experiments after specific learning events such as 
socialization or 
restraint stress have been reported to affect receptor density.
For instance, D2 receptor binding density in striatum is increased for socially 
dominant monkeys and reduced for subordinate monkeys 
\cite{MorganDetal2002} and induction of overexpression of D2 receptors in 
nucleus accumbens by genetic 
transfer in rats correlates with an increased resilience to alcohol 
addiction \cite{ThanosPKetal2001}.
Alterations in density of muscarinic receptors 
in neocortex and amygdala have been observed as the result of training
in an inhibitory avoidance task \cite{vanderZeeEAetal94,RoozendaalBetal97}.

Receptor density can be indirectly assessed by the binding capacity of 
receptors to radioligands (for membrane receptors) or the amount
of mRNA for a receptor (which comprises both internalized and membrane-bound 
receptor protein). With brain imaging (PET/SPECT), radioligand binding can even be 
measured in the living human brain \cite{LaruelleMetal97,Verhoeff99}.

Similarly to short-term desensitization, receptor density 
tends to increase at low concentrations of agonist and 
decrease at high levels of stimulation.
For instance, ongoing agonist stimulation ($>$ 4h) has been shown 
to result in long-term 
loss of membrane-bound receptor density for the dopamine D1 receptor in 
cultured cells (up to -50\% of control with 
$t_{1/2}$= 8h, \cite{SidhuAetal99}).
This basic homeostatic regulation is however embedded in a system of further
regulatory factors, which allow storage and permanence of receptor localization
and density.
Thus, long-term shifts in receptor density and efficacy may also be the 
result of specific brief co-ordinated events that affect relevant parameters in 
a sustained fashion.
In this sense, long-term plasticity in receptor density may have a 
memorization function that goes beyond adaptivity to current agonist 
concentrations.

Another question is the relationship between receptor density and functional
receptor sensitivity. Often, receptor density is directly parallel 
to functional receptor sensitivity as measured by effects on membrane 
excitability or synaptic transmission.
Thus agonist depletion produces both an increase in receptor 
density and an increase in electrophysiological responsiveness 
e.g. of beta-adrenergic receptors in hippocampus and cortex
\cite{ZahniserNRetal86} or for opioid receptors after 
morphine withdrawal \cite{MoisesSmith89}.

But increased sensitivity of a receptor may occur without an increase of
receptor density. 
The pharmacodynamic response to agonist occurs in proportion 
to the quantity of the ternary complex agonist-receptor-G-protein, 
not just receptor protein abundance per se, or even membrane-bound receptor 
protein abundance \cite{BursteinESetal97,SamamaPetal93}.
Receptors can exist both in a state where they are coupled to GTP-free 
G-proteins (high affinity to agonists), or they may exist without 
the effector molecule G-protein coupled to it, the low affinity state
\cite{CummingPetal2002}.
The supply of G-proteins is restricted in cells, such that 
there is competition for receptors to achieve a high affinity state.
Experimental evidence has indicated that there are shifts in 
affinity of D2 receptors in nucleus accumbens after exposure to amphetamine, 
which can explain increases of D2-receptor mediated 
behavior, even though the total receptor density remains unaltered
\cite{SeemanPetal2002}. In this case, D2 receptor supersensitivity 
is expressed by an increase in high affinity (G-protein coupled) receptor
sites \cite{SunWetal2003}.

This form of plasticity has certain implications. If we assume that 
receptor localization on the cell membrane (i.e. at synaptic sites) 
is relevant for neural transmission, changes in affinity can 
modulate transmission without affecting receptor localization.
This may be important for the retention of the functional properties of 
the modulated system.

Finally, long-term alterations of receptor sensitivity 
may be expressed 
by alterations in intracellular pathways, such as a permanent upregulation 
of the cAMP-pathway \cite{NestlerAghajanian97}.
In this case the shift in sensitivity is not specific to the type of 
receptor but to the pathway being modulated, which is usually connected 
to a number of different receptors.
Thus there is a third process available to effect long-term changes in 
receptor efficacy. 

Detailed experiments on dopamine receptor sensitivity have been conducted in 
slices of rat brain after exposure to cocaine or amphetamine. 

Sensitivity is here usually assessed by response threshold to different 
concentrations of agonist.
For nucleus 
accumbens in cocaine-sensitized animals, 
a much lower dose of dopamine ($20 \mu M$) elicits a electrophysiological 
response in D1 receptors than is required in control animals 
($75 \mu M$, \cite{BeurrierMalenka2002}).
The effect of the higher dose ($75 \mu M$) is the same in the 
supersensitive and normal system
\cite{BeurrierMalenka2002}. Thus there is a leftward shift in dose-dependence 
which is compatible with a change in the $I_{1}$ parameter for receptor
efficacy (see eq. \ref{eq-final} above).

Alternatively, the normal response 
to NM modulation may become
replaced by an ongoing "chronic" response and the cell becomes subsensitive 
or ceases to be  
responsive to agonist stimulation. This form of whole-cell plasticity is
not dependent on neuromodulator receptors, but consists of a shift in 
the ion channel distribution and density that defines membrane excitability
(e.g. \cite{ZhangXFetal98,HalterJAetal95}).
For instance, in cocaine-sensitized animals, some cells constitutively
suppress the N- and P-type calcium channels that are normally suppressed 
by D1 receptor activation \cite{ZhangXFetal2002}. The effect is a loss of 
NM responsivity on this parameter.




\subsection{NM effects on neural transmission}
The neural response is a product of both 
agonist concentration and the current receptor efficacy.
Previous computational models that attempted to assess the function of 
neuromodulation on the basis of variable agonist concentrations
\cite{SchultzWetal97,Schultz2002}, assuming that
receptor response would be uniform in space and time, may have failed to factor
in the real receptor variability, which, as we have seen, is highly regulated
and therefore likely to be of functional importance rather than only a 
source of unreliability and error ('noise').

Generally, neuromodulators have the potential to affect both synaptic 
transmission and intrinsic (whole-neuron) membrane properties.
The synaptic effects of the D1 receptor, and NM receptors in general, 
concern the regulation of transmitter release by presynaptic receptors 
and the regulation of NMDA-mediated glutamatergic transmission by 
postsynaptic receptors \cite{VittenIsaacson2001}.

Neuromodulators also have diffuse effects on the membrane 
potential of the neuron, mediated by alterations in ion channel currents 
from receptors on the dendrite and soma in postsynaptic and non-synaptic 
positions \cite{MarderEetal96,GoldmanMSetal2001}.
The significance of whole-neuron modulation, however, is considerably less 
well investigated as a source of memorization and learning.

There is a consensus that both presynaptic D1 
\cite{ChenXetal99,StenkampKetal98,GaoWJetal2001,Law-ThoDetal94,HarveyLacey96,UrbanNNetal2002,BehrJetal2000}
and D2 receptors 
\cite{HsuKSetal95,HsuKS96,EnnisMetal2001,KogaMomiyama2000,PisaniAetal2000}
depress the amplitude of evoked EPSP's in a number of different tissues 
with some debate as 
to whether the frequency of spontaneous EPSP's is similarly 
reduced \cite{NicolaSMetal96,BehrJetal2000}
or actually increased by presynaptic D1 receptors \cite{WangZetal2002}.
A similar effect has been observed for the regulation of GABA release: 
both D1 and D2 receptors reduce evoked IPSP's 
\cite{MomiyamaSim96,FedericiMetal2002,MomiyamaKoga2001,Gonzalez-IslasHablitz2001,ShenJohnson2000,DelgadoAetal2000,NicolaMalenka97}.
In postsynaptic 
position dopamine D1 receptors enhance NMDA transmission, by increasing 
peak conductance and lowering the threshold voltage for NMDA receptor 
activation \cite{WangODonnell2001,SeamansJKetal2001,Flores-HernandezJ1etal2002,CepedaCetal93}.
There may also be an effect of dopamine D1 receptors on AMPA receptors via 
phosphorylation of GluR1 subunits at $Ser^{845}$, mediated by inhibition 
of PP1 and PKA (via DARPP-32, \cite{SnyderGLetal2000})
in striatal neurons.
This conformational 
change increases the channel open time probability (in contrast to
phosphorylation 
at $Ser^{831}$ by PKC 
and caMKII, which increases AMPA channel conductance.)
Thus dopamine may also be able to enhance peak AMPA current 
in specific dopamine D1 receptor rich areas of the brain.

On the neuronal level, the effect of neuromodulators on signal transmission 
is expressed by altering membrane excitability.
However, it has been difficult to assess these effects precisely with the 
help of {\it in vitro} slices, and there is some disagreement concerning 
the effects of the dopamine D1 receptor.
%
The D1 receptor enhances or reduces the contribution of a number of ion 
channels, such as high-voltage activated calcium channels 
(L-type calcium channels are enhanced in neostriatal spiny 
neurons \cite{Hernandez-LopezSetal97}, N- and P-type calcium channels are 
blocked \cite{SurmeierDJetal95}).
It also affects sodium \cite{MauriceNetal2001} and potassium channels
\cite{NicolaSMetal2000}.

%
The effects on neuronal firing patterns are obviously complex, since they 
depend on membrane voltage, the distribution and frequency 
of different ion channels and other events affecting ion channel currents. 
%
%
%
%
%
%

\subsection{Synapse- and cell-specific receptor regulation}
We have seen that there is a significant body of evidence
showing that NM receptors undergo ex\-per\-ien\-ce-dependent long-term plasticity.

These results focus on the response of a {\it population} of neurons by 
examining a few selected neurons with the most pronounced alteration of 
response.
Results are usually not reported with an emphasis on 
{\it cell-specific} variability.
Nonetheless, there is often considerable heterogeneity with 
respect to reactivity to NM's within a population of neurons in a slice 
(J. Seamans, pers.  comm., for dopamine D1 and D2 receptors in 
deep-layer prefrontal cortical cells) with 'high responders' and 
'low responders' within a neuronal population.
Visual inspection also shows a varying number of fluorescent-labeled 
receptors on cultured cells, e.g. transfected with D2 receptors 
\cite{RayportSulzer95}, which is understood as an individual 
variability of receptor distribution and density (S. Rayport, pers. comm.).
Cellular binding patterns for dopamine receptors are different 
for different neurons also in striatal slices 
\cite{FisherRSetal94},\cite{ArianoSibley94},\cite{ArianoMAetal92}. 
New techniques of laser capture microscopy may
be able to give clear answers on the question of cell-specificity 
(M. Ariano, pers.comm).

There are also results indicating that receptor 
density may vary in 'patches' of (subcortical) tissue or microcolumns 
in cortical tissue \cite{vanderZeeEAetal95}. 
For instance, fear conditioning influences muscarinic receptor 
density differentially in different regions of the amygdala 
\cite{RoozendaalBetal97}, \cite{vanderZeeEAetal97}. 

Synapse-specific regulation has been shown conclusively in the short-term
\cite{ScottLetal2002}. NM receptors in synaptic positions are also
anchored by scaffolding and anchoring proteins, which indicates that 
positioning of membrane receptors at a specific site is relevant to NM 
receptor action \cite{LinRetal2001,BindaAVetal2002}. Thus, the 
basic mechanisms for synaptic long-term plasticity exist. Very recently, 
\cite{ColginLLetal2003} have actually 
shown direct synaptic plasticity at cholinergic synapses in hippocampus.

\subsection{Functional consequences of long-term plasticity}

We have argued that the effects of 
neuromodulation on neural transmission are not sufficiently described 
by the fluctuations in the concentration of agonist, but require 
an analysis of receptor sensitivity as well.
This means that the responsivity of the cell as expressed 
by receptor density and efficacy will determine the net effect on 
neural transmission.

The consequences of this regulation are somewhat different, whether we 
look at this from the perspective of a homogeneous shift in sensitivity 
in a particular brain area, or whether we assume cell-specific effects 
to be present under physiological conditions.

Thus
the plasticity located in the receptor and its internal transduction 
pathway could be {\it functionally} significant at the level of the 
individual cell and the synapse. This could be the case for both short-term and 
long-term plasticity.

\begin{figure}[htb]
\noindent
\begin{center}
\begin{minipage}[t]{0.48\textwidth}
\noindent{\bf A}
\includegraphics[height=3cm,width=0.9\textwidth]{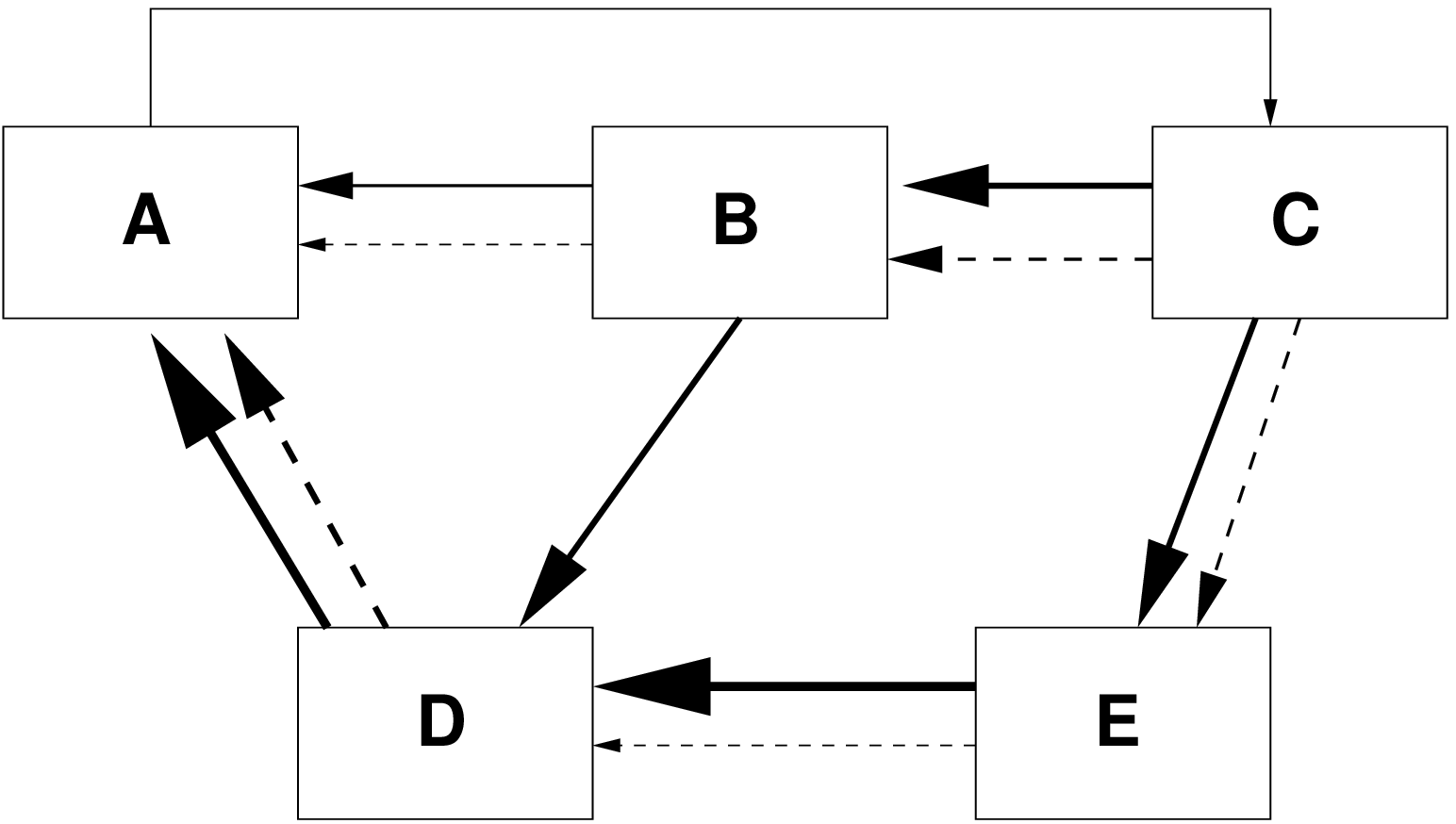}
\end{minipage}
\hfill
\begin{minipage}[t]{0.48\textwidth}
\noindent{\bf B}
\includegraphics[height=3cm,width=0.9\textwidth]{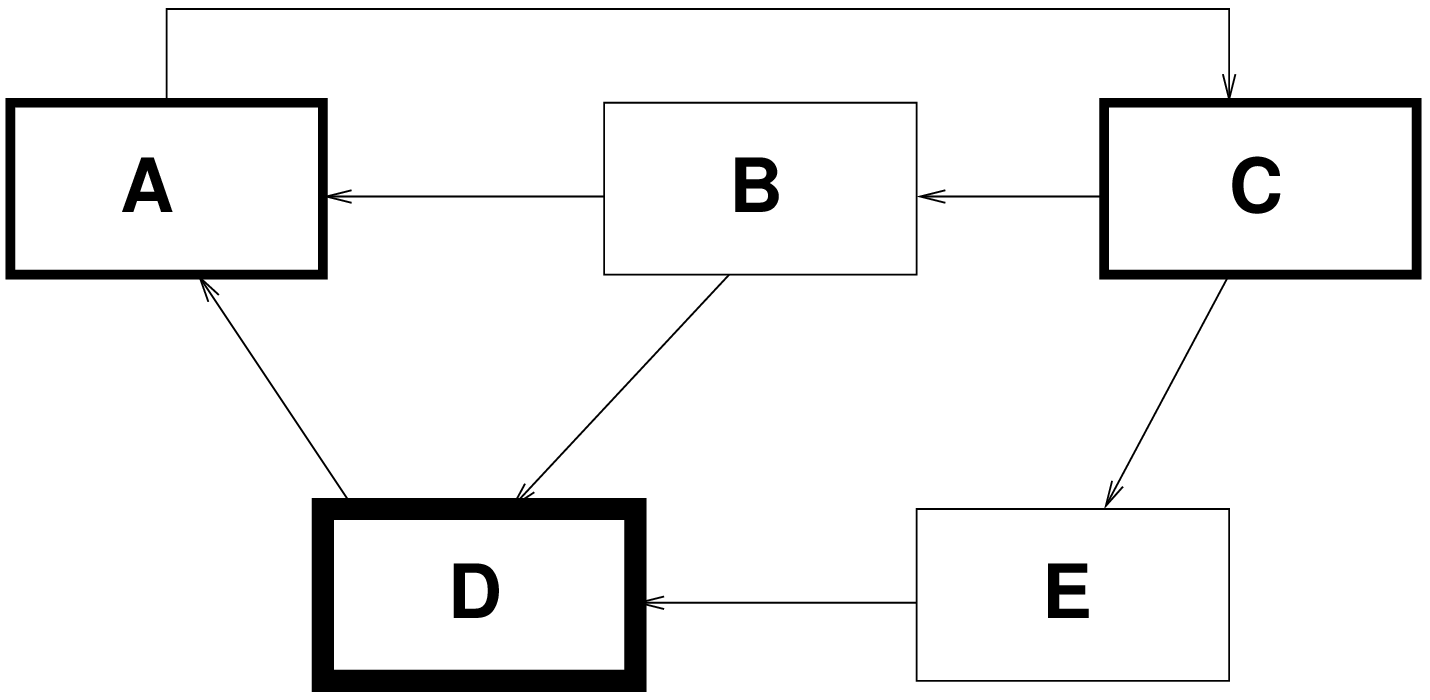}
\end{minipage}
\end{center}
\caption{Two expanded views on information storage in neural networks.
A: Processing units are linked by connections of differing strengths. 
Overloading and switching allows fast modification of synaptic strengths.
B: Processing units differ by their activation of transfer function.
Neurons with different membrane properties interact in a 'heterogeneous'
network.}
\label{syn-whole}
\end{figure}

Long-term plasticity in the responsivity of neurons and synapses to 
a neuromodulatory signal 
implies a mechanism of {\it information
storage} that is expressed by the placement of NM receptors leading to 
an individual cell signature.
Thus, NM receptor plasticity adds a layer of storage 
capacity
to neurons and synapses which allows memorization other than through 
Hebbian processes (see Fig. \ref{syn-whole}).
This hypothesis is considerably different 
from the widespread assumption that neuromodulation influences a global 
parameter for neural transmission which is uniform for all cells and synapses
in a targeted brain region \cite{Schultz2002,Koch99}.


Synapse-specific regulation of NM receptor efficacy and distribution specifically 
adds an important dimension to traditional accounts of long-term potentiation 
(LTP) and long-term depression (LTD), which focus on the contribution of the 
glutamatergic pathway.
It introduces additional variables which determine the magnitude of change of 
glutamate synaptic transmission when presynaptic NM receptors are 
activated and postsynaptic receptors interact with NMDA, AMPA or GABA$_A$ 
receptors. A NM signal-gated change in synaptic "weight" reflects a 
synapse-specific fixed parameter (namely the presence, absence or magnitude 
of a NM receptor mediated effect). The presence of a strong, 
phasic increase of NM concentration will then result in 
"fast synaptic switching" (Fig.~\ref{syn-whole} A). This means that synaptic 
connections can be quickly and reversibly set to a new weight, without 
interfering with any long-term properties of the glutamate transmission.
Technically, this corresponds to "overloading" of synaptic connections 
with several possible synaptic weights each of which can be switched on by 
activation of appropriate NM receptors.
Such mechanisms have occasionally been suggested 
in order to add to the known capabilities of neural networks
\cite{KonenWetal94,Schmidhuber92}.
In \cite{SchelerSchumann2003} it was shown that template switching by
neuromodulator control of presynaptic release 
is particularly amenable to the idea of state-dependent task performance, 
which is a major biological function of neuromodulation.

\subsection{Implications for brain plasticity}

The idea that neuromodulation contributes to long-term potentiation and 
long-term depression, influencing the magnitude of change in glutamatergic 
transmission, has been around for a long time. A number of experimental 
results have been obtained that support a measurable difference for
LTP/LTD in the presence or absence of high levels of neuromodulators, 
such as dopamine \cite{Manahan-VaughanKulla2003,BlondOetal2002}.
In this context, the theoretical concept of 'metaplasticity' has been 
developed 
\cite{AbrahamBear96}, cf. \cite{JayTM2003}. This concept assumes a level of 
regulation for 
glutamatergic plasticity that is not dependent on Hebbian associativity of
pre- and postsynaptic firing ('meta'-level). Another idea that forms the basis 
of a number of computational accounts of dopamine-related plasticity
\cite{WickensKoetter94,Schultz2002} is 'three-factor Hebbian 
plasticity', i.e. regulation of glutamatergic signalling strength that 
requires a 'third factor', namely dopamine, in addition to concurrence 
of presynaptic and postsynaptic events.
Both approaches, however, do not challenge the idea of glutamatergic 
signalling as a 'final common pathway' 
for different sources of plasticity and the changes in glutamatergic 
signalling as the substrate for learning.
 
In contrast to that, we have aimed to show that there is a significant 
motivation from the perspective of molecular biology that receptor 
plasticity is an ubiquitous phenomenon which may have become recruited 
for learning and memorization for a number of different neurochemicals.
We have also reported here that there is convincing electrophysiological 
evidence that shifts in receptor sensitivity do occur in the long-term,
even though specificity for individual cells and synapses is 
not well proven. The exact relationship of this form of plasticity to 
physiological brain adaptivity is at present virtually unknown.



The effects of NM receptor activation on {\it intrinsic properties} of the 
neuron - expressed by membrane excitability and ion channel activation -
add a significant dimension to brain plasticity. 
The most prevalent view of neural plasticity as altering the strength 
of connections between neurons is changed 
considerably, when we accept whole-neuron adaptive plasticity
(Fig.~\ref{syn-whole} B).

This role of 
whole-neuron plasticity within network processing 
has occasionally been 
explored from a theoretical perspective 
(cf. \cite{GoldmanMSetal2001,Magee2003}).
The alteration of membrane properties due to neuromodulation or a 
specific composition of ion channels induces a 'filter' on signal transmission
that may affect gain modulation \cite{SalinasSejnowski2001} or short-term 
retention 
of spike input patterns \cite{EgorovAVetal2002}.
A significant difference in receptor density and sensitivity will 
affect the efficacy of the intrinsic 'filter' for each individual neuron.

NM receptor regulation is specifically interesting for its property of
'conditional' plasticity. This means that differences in parameter
setting are greatly enhanced when a sufficient amount of 
agonist is present to engage NM receptors. 
The functional implications of NM receptor plasticity will 
be virtually non-existent or very strong depending on 
fluctuations in agonist concentration or the presence of an agonist
signal.

\section{Conclusion}

Even though the relevance of the biological processes underlying G-protein 
coupled receptor regulation to addiction  
research and psychopharmacology is frequently asserted 
\cite{NestlerAghajanian97}, we still do not have a good understanding of the physiological 
function of this form of plasticity.

The simplest theory would assert that receptor regulation is 
essentially a homeostatic control mechanism to counteract the significant 
fluctuations in agonist availability.
In this scenario, the goal of receptor regulation is to ensure a target 
range of NM effectiveness. 

Certainly that is a very important function of receptor regulation from a 
metabolic perspective. But the presence of multiple, nested feedback loops 
in a complex, highly regulated system 
suggests the presence of multistable solutions. 
The presence of receptor anchoring adds the necessary stability to transform 
transient fluctuations at synaptic sites into permanent values. 

In this paper we have focussed on the hypothesis :

\begin{description}
\item[\rm (a)] that neural information processing is influenced by the 
{\it combination} of the neurochemical signal (the agonist concentration) 
and the receptor response (the receptor efficacy or sensitivity),
\item[\rm (b)] that receptor efficacy is regulated on the level of the synapse 
and the neuron, and
\item[\rm (c)] that long-term plasticity of NM receptors is functionally 
significant in {\it information storage}.
\end{description}

The regulation of agonist concentration is of course another important 
factor in understanding neuromodulation. Even though short-term  changes 
are mainly a result of firing of the producer cells, transporter 
availability is another major factor that undergoes a form 
of functional plasticity 
\cite{LiuYetal99,BlakelyBauman2000,LetchworthSRetal2001}.


Other questions that require experimental analysis are the conditions 
that trigger the transition from short-term to long-term plasticity and 
the behavioral paradigms that influence long-term receptor density and 
placement. 
Certainly, further experimental evidence is required to explore the validity 
of this hypothesis.

Another important conclusion of this work is that 
the role that NMDA activation, calcium influx and enhanced 
protein synthesis play in a large number of behavioral learning experiments 
may have to be re-assessed without automatic reference to the 'dogma' of
glutamatergic synaptic plasticity as the neural substrate of learning and 
memory.
In contrast, NMDA- and calcium- related induction could point to a common, 
integrated perspective on neural plasticity, encompassing 
glutamatergic/GABAergic transmission, neuromodulation and the regulation of 
internal cell processes.

We conclude that we need to pursue integrated models of neural and 
synaptic plasticity, which combine AMPA and glutamate related 
plasticity and NM related plasticity into a single model.
Fundamentally new theoretical abstractions need to be developed that can 
provide a guideline in the experimental testing of their implications.
Essentially we will have to explore the agonist-dependence of NM receptor
sensitivity and the concept of conditional plasticity as the basis of 
"learning rules" for neuromodulation (e.g., \cite{FransenEetal2002}).

Non-traditional sources of plasticity that may contribute to models 
of memory and learning are not restricted to plasticity in 
NM receptor activity.
They include long-term alterations of the distribution of ion channels,  
morphological alterations in spine density and dendritic branching, 
and levels of gene expression for a number of proteins affecting intracellular
pathways.

The fundamental dogma of Hebbian plasticity - 
associative strengthening of the synaptic connections that mediate fast 
neural transmission as neural substrate for learning - may have to be 
reconsidered.

The modification of synaptic connections has undoubtedly  
a major role in experience-dependent plasticity  - but  
the idea of associative strengthening based on the co-occurence 
(or, in more modern views, the precise timing) of 
pre- and postsynaptic activity may have to be embedded in a more comprehensive
view, involving NM induced switching, overloading of synaptic values, or
adaptive regulation of coupling strength. In general, the processes that 
govern cross-signaling in transmitter release are not sufficiently understood,
but undoubtedly of importance in understanding synaptic plasticity, as 
well as the interaction of NM receptors with glutamate and GABA receptors 
in the postsynaptic domain.

But besides that, we may have to further investigate the 
role of whole-neuron plasticity in learning and memorization. 
This involves the storage of experience by 
cellular parameters and performing read-out by alteration of membrane 
properties that affect the transfer function.  

In the light of the major potential sources of cellular plasticity uncovered 
by molecular biology, it may well be that the most significant models 
on the neural substrate of behavioral learning 
and memorization are yet ahead of us.


\bibliographystyle{elsart-num}

\newpage


\begin{thebibliography}{100}
\expandafter\ifx\csname url\endcsname\relax
  \def\url#1{\texttt{#1}}\fi
\expandafter\ifx\csname urlprefix\endcsname\relax\def\urlprefix{URL }\fi

\bibitem{HenryWhite95}
D.~J. Henry, F.~J. White, {The persistence of behavioral sensitization to
  cocaine parallels enhanced inhibition of nucleus accumbens neurons.}, J
  Neurosci 15 (1995) 6287--99.

\bibitem{WangWetal96}
W.~Wang, K.~H. Hahn, J.~F. Bishop, D.~Q. Gao, P.~A. Jose, M.~M. Mouradian,
  {Up-regulation of D3 dopamine receptor mRNA by neuroleptics.}, Synapse 23
  (1996) 232--5.

\bibitem{ScheggiSetal2002}
S.~Scheggi, B.~Leggio, F.~Masi, S.~Grappi, C.~Gambarana, G.~Nanni, R.~Rauggi,
  M.~G. De~Montis, {Selective modifications in the nucleus accumbens of
  dopamine synaptic transmission in rats exposed to chronic stress.}, J
  Neurochem 83 (2002) 895--903.

\bibitem{OrtegaAetal96}
A.~Ortega, M.~A. del Guante, R.~A. Prado-Alcala, V.~Aleman, {Changes in rat
  brain muscarinic receptors after inhibitory avoidance learning.}, Life Sci 58
  (1996) 799--809.

\bibitem{vanderZeeLuiten99}
E.~A. van~der Zee, P.~G. Luiten, {Muscarinic acetylcholine receptors in the
  hippocampus, neocortex and amygdala: a review of immunocytochemical
  localization in relation to learning and memory.}, Prog Neurobiol 58 (1999)
  409--71.

\bibitem{vanderZeeEAetal94}
E.~A. Van~der Zee, B.~R. Douma, B.~Bohus, P.~G. Luiten, {Passive avoidance
  training induces enhanced levels of immunoreactivity for muscarinic
  acetylcholine receptor and coexpressed PKC gamma and MAP-2 in rat cortical
  neurons.}, Cereb Cortex 4 (1994) 376--90.

\bibitem{MorganDetal2002}
D.~Morgan, K.~A. Grant, H.~D. Gage, R.~H. Mach, J.~R. Kaplan, O.~Prioleau,
  S.~H. Nader, N.~Buchheimer, R.~L. Ehrenkaufer, M.~A. Nader, {Social dominance
  in monkeys: dopamine D2 receptors and cocaine self-administration.}, Nat
  Neurosci 5 (2002) 169--74.

\bibitem{LanfumeyLetal99}
L.~Lanfumey, M.~C. Pardon, N.~Laaris, C.~Joubert, N.~Hanoun, M.~Hamon,
  C.~Cohen-Salmon, {5-HT1A autoreceptor desensitization by chronic ultramild
  stress in mice.}, Neuroreport 10 (1999) 3369--74.

\bibitem{JakabGoldman-Rakic2000}
R.~L. Jakab, P.~S. Goldman-Rakic, {Segregation of serotonin 5-HT2A and 5-HT3
  receptors in inhibitory circuits of the primate cerebral cortex.}, J Comp
  Neurol 417 (2000) 337--48.

\bibitem{LeMoineGaspar98}
C.~Le~Moine, P.~Gaspar, {Subpopulations of cortical GABAergic interneurons
  differ by their expression of D1 and D2 dopamine receptor subtypes.}, Brain
  Res Mol Brain Res 58 (1998) 231--6.

\bibitem{SurmeierDJetal96}
D.~J. Surmeier, W.~J. Song, Z.~Yan, {Coordinated expression of dopamine
  receptors in neostriatal medium spiny neurons.}, J Neurosci 16 (1996)
  6579--91.

\bibitem{YungBolam2000}
K.~K. Yung, J.~P. Bolam, {Localization of dopamine D1 and D2 receptors in the
  rat neostriatum: synaptic interaction with glutamate- and GABA-containing
  axonal terminals.}, Synapse 38 (2000) 413--20.

\bibitem{MengSZetal99}
S.~Z. Meng, Y.~Ozawa, M.~Itoh, S.~Takashima, {Developmental and age-related
  changes of dopamine transporter, and dopamine D1 and D2 receptors in human
  basal ganglia.}, Brain Res 843 (1999) 136--44.

\bibitem{GurevichEVetal2001}
E.~V. Gurevich, R.~T. Robertson, J.~N. Joyce, {Thalamo-cortical afferents
  control transient expression of the dopamine D(3) receptor in the rat
  somatosensory cortex.}, Cereb Cortex 11 (2001) 691--701.

\bibitem{MontagueDMetal99}
D.~M. Montague, C.~P. Lawler, R.~B. Mailman, J.~H. Gilmore, {Developmental
  regulation of the dopamine D1 receptor in human caudate and putamen.},
  Neuropsychopharmacology 21 (1999) 641--9.

\bibitem{TeicherMHetal95}
M.~H. Teicher, S.~L. Andersen, J.~C.~J. Hostetter, {Evidence for dopamine
  receptor pruning between adolescence and adulthood in striatum but not
  nucleus accumbens.}, Brain Res Dev Brain Res 89 (1995) 167--72.

\bibitem{PowerJMetal2002}
J.~M. Power, W.~W. Wu, E.~Sametsky, M.~M. Oh, J.~F. Disterhoft, {Age-related
  enhancement of the slow outward calcium-activated potassium current in
  hippocampal CA1 pyramidal neurons in vitro.}, J Neurosci 22 (2002) 7234--43.

\bibitem{KrupnickBenovic98}
J.~G. Krupnick, J.~L. Benovic, {The role of receptor kinases and arrestins in G
  protein-coupled receptor regulation.}, Annu Rev Pharmacol Toxicol 38 (1998)
  289--319.

\bibitem{HausdorffWPetal90}
W.~P. Hausdorff, M.~J. Lohse, M.~Bouvier, S.~B. Liggett, M.~G. Caron, R.~J.
  Lefkowitz, {Two kinases mediate agonist-dependent phosphorylation and
  desensitization of the beta 2-adrenergic receptor.}, Symp Soc Exp Biol 44
  (1990) 225--40.

\bibitem{MomiyamaKoga2001}
T.~Momiyama, E.~Koga, {Dopamine D(2)-like receptors selectively block N-type
  Ca(2+) channels to reduce GABA release onto rat striatal cholinergic
  interneurones.}, J Physiol 533 (2001) 479--92.

\bibitem{NicolaSMetal2000}
S.~M. Nicola, J.~Surmeier, R.~C. Malenka, {Dopaminergic modulation of neuronal
  excitability in the striatum and nucleus accumbens.}, Annu Rev Neurosci 23
  (2000) 185--215.

\bibitem{Hernandez-LopezSetal97}
S.~Hernandez-Lopez, J.~Bargas, D.~J. Surmeier, A.~Reyes, E.~Galarraga, {D1
  receptor activation enhances evoked discharge in neostriatal medium spiny
  neurons by modulating an L-type Ca2+ conductance.}, J Neurosci 17 (1997)
  3334--42.

\bibitem{SurmeierDJetal95}
D.~J. Surmeier, J.~Bargas, H.~C.~J. Hemmings, A.~C. Nairn, P.~Greengard,
  {Modulation of calcium currents by a D1 dopaminergic protein
  kinase/phosphatase cascade in rat neostriatal neurons.}, Neuron 14 (1995)
  385--97.

\bibitem{CantrellARetal99}
A.~R. Cantrell, T.~Scheuer, W.~A. Catterall, {Voltage-dependent neuromodulation
  of Na+ channels by D1-like dopamine receptors in rat hippocampal neurons.}, J
  Neurosci 19 (1999) 5301--10.

\bibitem{MauriceNetal2001}
N.~Maurice, T.~Tkatch, M.~Meisler, L.~K. Sprunger, D.~J. Surmeier, {D1/D5
  dopamine receptor activation differentially modulates rapidly inactivating
  and persistent sodium currents in prefrontal cortex pyramidal neurons.}, J
  Neurosci 21 (2001) 2268--77.

\bibitem{WangODonnell2001}
J.~Wang, P.~O'Donnell, {D(1) dopamine receptors potentiate nmda-mediated
  excitability increase in layer V prefrontal cortical pyramidal neurons.},
  Cereb Cortex 11 (2001) 452--62.

\bibitem{Flores-HernandezJ1etal2002}
J.~Flores-Hernandez, C.~Cepeda, E.~Hernandez-Echeagaray, C.~R. Calvert, E.~S.
  Jokel, A.~A. Fienberg, P.~Greengard, M.~S. Levine, {Dopamine enhancement of
  NMDA currents in dissociated medium-sized striatal neurons: role of D1
  receptors and DARPP-32.}, J Neurophysiol 88 (2002) 3010--20.

\bibitem{KassackMUetal2002}
M.~U. Kassack, B.~Hofgen, J.~Lehmann, N.~Eckstein, J.~M. Quillan, W.~Sadee,
  {Functional screening of G protein-coupled receptors by measuring
  intracellular calcium with a fluorescence microplate reader.}, J Biomol
  Screen 7 (2002) 233--46.

\bibitem{ZhuangXetal2000}
X.~Zhuang, L.~Belluscio, R.~Hen, {G$_{olf\alpha}$ mediates dopamine D1 receptor
  signaling.}, J Neurosci RC91 (2000) 1--5.

\bibitem{FergusonSS2001}
S.~S. Ferguson, {Evolving concepts in G protein-coupled receptor endocytosis:
  the role in receptor desensitization and signaling.}, Pharmacol Rev 53 (2001)
  1--24.

\bibitem{CarmanBenovic98}
C.~V. Carman, J.~L. Benovic, {G-protein-coupled receptors: turn-ons and
  turn-offs.}, Curr Opin Neurobiol 8 (1998) 335--44.

\bibitem{KabbaniNetal2002}
N.~Kabbani, L.~Negyessy, R.~Lin, P.~Goldman-Rakic, R.~Levenson, {Interaction
  with neuronal calcium sensor NCS-1 mediates desensitization of the D2
  dopamine receptor.}, J Neurosci 22 (2002) 8476--86.

\bibitem{LohseMJetal90}
M.~J. Lohse, J.~L. Benovic, J.~Codina, M.~G. Caron, R.~J. Lefkowitz,
  {beta-Arrestin: a protein that regulates beta-adrenergic receptor function.},
  Science 248 (1990) 1547--50.

\bibitem{JeongIkeda2000}
S.~W. Jeong, S.~R. Ikeda, {Endogenous regulator of G-protein signaling proteins
  modify N-type calcium channel modulation in rat sympathetic neurons.}, J
  Neurosci 20 (2000) 4489--96.

\bibitem{vonZastrowMostov2001}
M.~von Zastrow, K.~Mostov, {Signal transduction. A new thread in an intricate
  web.}, Science 294 (2001) 1845--7.

\bibitem{ZhengBetal2001}
B.~Zheng, Y.~C. Ma, R.~S. Ostrom, C.~Lavoie, G.~N. Gill, P.~A. Insel, X.~Y.
  Huang, M.~G. Farquhar, {RGS-PX1, a GAP for GalphaS and sorting nexin in
  vesicular trafficking.}, Science 294 (2001) 1939--42.

\bibitem{GeurtsMetal2002}
M.~Geurts, E.~Hermans, J.~M. Maloteaux, {Opposite modulation of regulators of G
  protein signalling-2 (RGS2) and RGS4 expression by dopamine receptors in the
  rat striatum.}, Neurosci Lett (2002) 146--50.

\bibitem{TaymansJMetal2003}
J.~M. Taymans, J.~E. Leysen, X.~Langlois, {Striatal gene expression of RGS2 and
  RGS4 is specifically mediated by dopamine D1 and D2 receptors: clues for RGS2
  and RGS4 function.}, J Neurochem (2003) 1118--1127.

\bibitem{JiangSibley99}
D.~Jiang, D.~R. Sibley, {Regulation of D(1) dopamine receptors with mutations
  of protein kinase phosphorylation sites: attenuation of the rate of
  agonist-induced desensitization.}, Mol Pharmacol 56 (1999) 675--83.

\bibitem{MasonJNetal2002}
J.~N. Mason, L.~B. Kozell, K.~A. Neve, {Regulation of dopamine D(1) receptor
  trafficking by protein kinase A-dependent phosphorylation.}, Mol Pharmacol 61
  (2002) 806--16.

\bibitem{GardnerBetal2001}
B.~Gardner, Z.~F. Liu, D.~Jiang, D.~R. Sibley, {The role of
  phosphorylation/dephosphorylation in agonist-induced desensitization of D1
  dopamine receptor function: evidence for a novel pathway for receptor
  dephosphorylation.}, Mol Pharmacol 59 (2001) 310--21.

\bibitem{VickeryvonZastrow99}
R.~G. Vickery, M.~von Zastrow, {Distinct dynamin-dependent and -independent
  mechanisms target structurally homologous dopamine receptors to different
  endocytic membranes.}, J Cell Biol 144 (1999) 31--43.

\bibitem{MorrisonKJetal96}
K.~J. Morrison, R.~H. Moore, N.~D. Carsrud, J.~Trial, E.~E. Millman, M.~Tuvim,
  R.~B. Clark, R.~Barber, B.~F. Dickey, B.~J. Knoll, {Repetitive endocytosis
  and recycling of the beta 2-adrenergic receptor during agonist-induced steady
  state redistribution.}, Mol Pharmacol 50 (1996) 692--9.

\bibitem{ChuangTTetal96}
T.~T. Chuang, L.~Paolucci, A.~De~Blasi, {Inhibition of G protein-coupled
  receptor kinase subtypes by Ca2+/calmodulin.}, J Biol Chem 271 (1996)
  28691--6.

\bibitem{HagaKetal97}
K.~Haga, H.~Tsuga, T.~Haga, {Ca2+-dependent inhibition of G protein-coupled
  receptor kinase 2 by calmodulin.}, Biochemistry 36 (1997) 1315--21.

\bibitem{ProninANetal97}
A.~N. Pronin, D.~K. Satpaev, V.~Z. Slepak, J.~L. Benovic, {Regulation of G
  protein-coupled receptor kinases by calmodulin and localization of the
  calmodulin binding domain.}, J Biol Chem 272 (1997) 18273--80.

\bibitem{KraselCetal2001}
C.~Krasel, S.~Dammeier, R.~Winstel, J.~Brockmann, H.~Mischak, M.~J. Lohse,
  {Phosphorylation of GRK2 by protein kinase C abolishes its inhibition by
  calmodulin.}, J Biol Chem 276 (2001) 1911--5.

\bibitem{BouvierMetal88}
M.~Bouvier, W.~P. Hausdorff, A.~De~Blasi, B.~F. O'Dowd, B.~K. Kobilka, M.~G.
  Caron, R.~J. Lefkowitz, {Removal of phosphorylation sites from the beta
  2-adrenergic receptor delays onset of agonist-promoted desensitization.},
  Nature 333 (1988) 370--3.

\bibitem{NgGYetal94}
G.~Y. Ng, B.~Mouillac, S.~R. George, M.~Caron, M.~Dennis, M.~Bouvier, B.~F.
  O'Dowd, {Desensitization, phosphorylation and palmitoylation of the human
  dopamine D1 receptor.}, Eur J Pharmacol 267 (1994) 7--19.

\bibitem{TiberiMetal96}
M.~Tiberi, S.~R. Nash, L.~Bertrand, R.~J. Lefkowitz, M.~G. Caron, {Differential
  regulation of dopamine D1A receptor responsiveness by various G
  protein-coupled receptor kinases.}, J Biol Chem 271 (1996) 3771--8.

\bibitem{SibleyDRetal98}
D.~R. Sibley, A.~L. Ventura, D.~Jiang, C.~Mak, {Regulation of the D1 dopamine
  receptor through cAMP-mediated pathways.}, Adv Pharmacol 42 (1998) 447--50.

\bibitem{VenturaSibley2000}
A.~L. Ventura, D.~R. Sibley, {Altered regulation of the D(1) dopamine receptor
  in mutant Chinese hamster ovary cells deficient in cyclic AMP-dependent
  protein kinase activity.}, J Pharmacol Exp Ther 293 (2000) 426--34.

\bibitem{BatesMDetal91}
M.~D. Bates, M.~G. Caron, J.~R. Raymond, {Desensitization of DA1 dopamine
  receptors coupled to adenylyl cyclase in opossum kidney cells.}, Am J Physiol
  260 (1991) F937--45.

\bibitem{BlackLEetal94}
L.~E. Black, E.~M. Smyk-Randall, D.~R. Sibley, {Cyclic AMP-mediated
  desensitization of D1 dopamine receptor-coupled adenylyl cyclase in NS20Y
  neuroblastoma cells.}, Mol Cell Neurosci 5 (1994) 567--75.

\bibitem{LewisMMetal98}
M.~M. Lewis, V.~J. Watts, C.~P. Lawler, D.~E. Nichols, R.~B. Mailman,
  {Homologous desensitization of the D1A dopamine receptor: efficacy in causing
  desensitization dissociates from both receptor occupancy and functional
  potency.}, J Pharmacol Exp Ther 286 (1998) 345--53.

\bibitem{BatesMDetal93}
M.~D. Bates, C.~L. Olsen, B.~N. Becker, F.~J. Albers, J.~P. Middleton, J.~G.
  Mulheron, S.~L. Jin, M.~Conti, J.~R. Raymond, {Elevation of cAMP is required
  for down-regulation, but not agonist-induced desensitization, of endogenous
  dopamine D1 receptors in opossum kidney cells. Studies in cells that stably
  express a rat cAMP phosphodiesterase (rPDE3) cDNA.}, J Biol Chem 268 (1993)
  14757--63.

\bibitem{NgGYetal95}
G.~Y. Ng, J.~Trogadis, J.~Stevens, M.~Bouvier, B.~F. O'Dowd, S.~R. George,
  {Agonist-induced desensitization of dopamine D1 receptor-stimulated adenylyl
  cyclase activity is temporally and biochemically separated from D1 receptor
  internalization.}, Proc Natl Acad Sci U S A 92 (1995) 10157--61.

\bibitem{NgGYetal97}
G.~Y. Ng, G.~Varghese, H.~T. Chung, J.~Trogadis, P.~Seeman, B.~F. O'Dowd, S.~R.
  George, {Resistance of the dopamine D2L receptor to desensitization
  accompanies the up-regulation of receptors on to the surface of Sf9 cells.},
  Endocrinology 138 (1997) 4199--206.

\bibitem{ChenLambert2000}
H.~Chen, N.~A. Lambert, {Endogeneous regulators of G protein signaling proteins
  regulate presynaptic inhibition at rat hippocampal synapses.}, Proc Natl Acad
  Sci U S A 97 (2000) 12810--12815.

\bibitem{JeongIkeda98}
S.~W. Jeong, S.~R. Ikeda, {G Protein $\alpha$ Subunit $G_{\alpha}z$ couples
  neurotransmitter receptors to ion channels in sympathetic neurons.}, Neuron
  21 (1998) 1201--1212.

\bibitem{SaitohOetal97}
O.~Saitoh, Y.~Kubo, Y.~Miyatani, T.~Asano, H.~Nakata, {RGS8 accelerates
  G-protein-mediated modulation of K+ currents.}, Nature 390 (1997) 525--529.

\bibitem{DoupnikCAetal97}
C.~A. Doupnik, N.~Davidson, H.~A. Lester, P.~Kofuji, {RGS proteins reconstitute
  the rapid gating kinetics of $G_{\beta\gamma}$-activated inwardly rectifying
  K+ channels.}, Proc Natl Acad Sci U S A 94 (1997) 10461--10466.

\bibitem{WatsonNetal96}
N.~Watson, M.~E. Linder, K.~M. Druey, J.~H. Kehrl, K.~J. Blumer, {RGS family
  members: GTPase-activating proteins for heterotrimeric G-protein
  alpha-subunits.}, Nature 383 (1996) 172--175.

\bibitem{TiberiCaron94}
M.~Tiberi, M.~G. Caron, {High agonist-independent activity is a distinguishing
  feature of the dopamine D1B receptor subtype.}, J Biol Chem 269 (1994)
  27925--31.

\bibitem{GrotewielSanders-Bush99}
M.~S. Grotewiel, E.~Sanders-Bush, {Differences in agonist-independent activity
  of 5-Ht2A and 5-HT2c receptors revealed by heterologous expression.}, Naunyn
  Schmiedebergs Arch Pharmacol 359 (1999) 21--7.

\bibitem{DemchyshynLLetal2000}
L.~L. Demchyshyn, F.~McConkey, H.~B. Niznik, {Dopamine D5 receptor agonist high
  affinity and constitutive activity profile conferred by carboxyl-terminal
  tail sequence.}, J Biol Chem 275 (2000) 23446--55.

\bibitem{LameyMetal2002}
M.~Lamey, M.~Thompson, G.~Varghese, H.~Chi, M.~Sawzdargo, S.~R. George, B.~F.
  O'Dowd, {Distinct residues in the carboxyl tail mediate agonist-induced
  desensitization and internalization of the human dopamine D1 receptor.}, J
  Biol Chem 277 (2002) 9415--21.

\bibitem{BhallaIyengar2001}
U.~S. Bhalla, R.~Iyengar, {Robustness of the bistable behavior of a biological
  signaling feedback loop.}, Chaos 11 (2001) 221--226.

\bibitem{Bhalla2002}
U.~S. Bhalla, {The chemical organization of signaling interactions.},
  Bioinformatics 18 (2002) 855--63.

\bibitem{WengGetal99}
G.~Weng, U.~S. Bhalla, R.~Iyengar, {Complexity in biological signaling
  systems.}, Science 284 (1999) 92--96.

\bibitem{MishraBhalla2002}
J.~Mishra, U.~S. Bhalla, {Simulations of inositol phosphate metabolism and its
  interaction with InsP(3)-mediated calcium release.}, Biophys J 83 (2002)
  1298--1316.

\bibitem{LeeEetal2003}
E.~Lee, A.~Salic, R.~Krueger, M.~W. Kirschner, {The Roles of APC and Axin
  Derived from Experimental and Theoretical Analysis of the Wnt Pathway.}, PLoS
  Biology 1 (2003) 116--132.

\bibitem{WileyHSetal2003}
H.~S. Wiley, S.~Y. Shvartsman, D.~A. Lauffenburger, {Computational modeling of
  the EGF-receptor system: a paradigm for systems biology.}, Trends in Cell
  Biology 13 (2003) 43--50.

\bibitem{Bhalla2003}
U.~S. Bhalla, {Understanding complex signaling networks through models and
  metaphors.}, Prog Biophys Mol Biol 81 (2003) 45--65.

\bibitem{TsaovonZastrow2000}
P.~Tsao, M.~von Zastrow, {Downregulation of G protein-coupled receptors.}, Curr
  Opin Neurobiol 10 (2000) 365--9.

\bibitem{CaiGetal2002}
G.~Cai, H.-Y. Wang, E.~Friedman, {Increased dopamine receptor signaling and
  dopamine receptor-G protein coupling in denervated striatum.}, J Pharmacol
  Exp Ther 302 (2002) 1105--12.

\bibitem{KimDSetal2000}
D.~S. Kim, M.~S. Szczypka, R.~D. Palmiter, {Dopamine-deficient mice are
  hypersensitive to dopamine receptor agonists.}, J Neurosci 20 (2000)
  4405--13.

\bibitem{UnterwaldEMetal94}
E.~M. Unterwald, A.~Ho, J.~M. Rubenfeld, M.~J. Kreek, {Time course of the
  development of behavioral sensitization and dopamine receptor up-regulation
  during binge cocaine administration.}, J Pharmacol Exp Ther 270 (1994)
  1387--96.

\bibitem{HenryDJetal98}
D.~J. Henry, X.~T. Hu, F.~J. White, {Adaptations in the mesoaccumbens dopamine
  system resulting from repeated administration of dopamine D1 and D2
  receptor-selective agonists: relevance to cocaine sensitization.},
  Psychopharmacology (Berl) 140 (1998) 233--42.

\bibitem{MayfieldRDetal92}
R.~D. Mayfield, G.~Larson, N.~R. Zahniser, {Cocaine-induced behavioral
  sensitization and D1 dopamine receptor function in rat nucleus accumbens and
  striatum.}, Brain Res 573 (1992) 331--5.

\bibitem{HuXTetal2002}
X.-T. Hu, T.~E. Koeltzow, D.~C. Cooper, G.~S. Robertson, F.~J. White,
  P.~Vezina, {Repeated ventral tegmental area amphetamine administration alters
  dopamine D1 receptor signaling in the nucleus accumbens.}, Synapse 45 (2002)
  159--70.

\bibitem{ChenJCetal99}
J.~C. Chen, H.~J. Su, L.~I. Huang, M.~M. Hsieh, {Reductions in binding and
  functions of D2 dopamine receptors in the rat ventral striatum during
  amphetamine sensitization.}, Life Sci 64 (1999) 343--54.

\bibitem{SeeREetal90}
R.~E. See, A.~W. Toga, G.~Ellison, { Autoradiographic analysis of regional
  alterations in brain receptors following chronic administration and
  withdrawal of typical and atypical neuroleptics in rats.}, J Neural Transm
  Gen Sect 82 (1990) 93--109.

\bibitem{TaraziFIetal97}
F.~Tarazi, W.~J. Florijn, I.~Creese, {Differential regulation of dopamine
  receptors after chronic typical and atypical antipsychotic drug treatment.},
  Neuroscience 78 (1997) 985--996.

\bibitem{LeeHetal99}
L.~H, F.~Tarazi, M.~Chakos, H.~Wu, M.~Redmond, J.~Alvir, B.~Kinon, R.~Bilder,
  I.~Creese, J.~Lieberman, {Effects of chronic treatment with typical and
  atypical antipsychotic drugs on the rat striatum.}, Life Sci 64 (1999)
  1595--1602.

\bibitem{ThanosPKetal2001}
P.~K. Thanos, N.~D. Volkow, P.~Freimuth, H.~Umegaki, H.~Ikari, G.~Roth, D.~K.
  Ingram, R.~Hitzemann, {Overexpression of dopamine D2 receptors reduces
  alcohol self-administration.}, J Neurochem 78 (2001) 1094--103.

\bibitem{RoozendaalBetal97}
B.~Roozendaal, E.~A. van~der Zee, R.~A. Hensbroek, H.~Maat, P.~G. Luiten, J.~M.
  Koolhaas, B.~Bohus, {Muscarinic acetylcholine receptor immunoreactivity in
  the amygdala--II. Fear-induced plasticity.}, Neuroscience 76 (1997) 75--83.

\bibitem{LaruelleMetal97}
M.~Laruelle, C.~D. D'Souza, R.~M. Baldwin, A.~Abi-Dargham, S.~J. Kanes, C.~L.
  Fingado, J.~P. Seibyl, S.~S. Zoghbi, M.~B. Bowers, P.~Jatlow, D.~S. Charney,
  R.~B. Innis, {Imaging D2 receptor occupancy by endogenous dopamine in
  humans.}, Neuropsychopharmacology 17 (1997) 162--74.

\bibitem{Verhoeff99}
N.~P. Verhoeff, {Radiotracer imaging of dopaminergic transmission in
  neuropsychiatric disorders.}, Psychopharmacology (Berl) 147 (1999) 217--49.

\bibitem{SidhuAetal99}
A.~Sidhu, B.~Olde, N.~Humblot, K.~Kimura, N.~Gardner, {Regulation of human D1
  dopamine receptor function and gene expression in SK-N-MC neuroblastoma
  cells.}, Neuroscience 91 (1999) 537--47.

\bibitem{ZahniserNRetal86}
N.~R. Zahniser, G.~R. Weiner, T.~Worth, K.~Philpott, R.~P. Yasuda, G.~Jonsson,
  T.~V. Dunwiddie, {DSP4-induced noradrenergic lesions increase beta-adrenergic
  receptors and hippocampal electrophysiological responsiveness.}, Pharmacol
  Biochem Behav 24 (1986) 1397--402.

\bibitem{MoisesSmith89}
H.~C. Moises, C.~B. Smith, {Electrophysiological responsiveness to
  isoproterenol in rat hippocampal slices correlates with changes in
  beta-adrenergic receptor density induced by chronic morphine treatment.},
  Brain Res 485 (1989) 67--78.

\bibitem{BursteinESetal97}
E.~S. Burstein, T.~A. Spalding, M.~R. Brann, {Pharmacology of muscarinic
  receptor subtypes constitutively activated by G proteins.}, Mol Pharmacol 51
  (1997) 312--9.

\bibitem{SamamaPetal93}
P.~Samama, S.~Cotecchia, T.~Costa, R.~J. Lefkowitz, {A mutation-induced
  activated state of the beta 2-adrenergic receptor. Extending the ternary
  complex model.}, J Biol Chem 268 (1993) 4625--36.

\bibitem{CummingPetal2002}
P.~Cumming, D.~F. Wong, R.~F. Dannals, N.~Gillings, J.~Hilton, U.~Scheffel,
  A.~Gjedde, {The competition between endogenous dopamine and radioligands for
  specific binding to dopamine receptors.}, Ann N Y Acad Sci 965 (2002)
  440--50.

\bibitem{SeemanPetal2002}
P.~Seeman, T.~Tallerico, F.~Ko, C.~Tenn, S.~Kapur, {Amphetamine-sensitized
  animals show a marked increase in dopamine D2 high receptors occupied by
  endogenous dopamine, even in the absence of acute challenges.}, Synapse 46
  (2002) 235--9.

\bibitem{SunWetal2003}
W.~Sun, N.~Ginovart, F.~Ko, P.~Seeman, S.~Kapur, {In vivo evidence for
  dopamine-mediated internalization of D2-receptors after amphetamine:
  differential findings with [3H]raclopride versus [3H]spiperone.}, Mol
  Pharmacol 63 (2003) 456--62.

\bibitem{NestlerAghajanian97}
E.~J. Nestler, G.~K. Aghajanian, {Molecular and cellular basis of addiction.},
  Science 278 (1997) 58--63.

\bibitem{BeurrierMalenka2002}
C.~Beurrier, R.~C. Malenka, {Enhanced inhibition of synaptic transmission by
  dopamine in the nucleus accumbens during behavioral sensitization to
  cocaine.}, J Neurosci 22 (2002) 5817--22.

\bibitem{ZhangXFetal98}
X.-F. Zhang, X.~T. Hu, F.~J. White, {Whole-cell plasticity in cocaine
  withdrawal: reduced sodium currents in nucleus accumbens neurons.}, J
  Neurosci 18 (1998) 488--98.

\bibitem{HalterJAetal95}
J.~A. Halter, J.~S. Carp, J.~R. Wolpaw, {Operantly conditioned motoneuron
  plasticity:possible role of sodium channels.}, J Neurophysiol 73 (1995)
  867--871.

\bibitem{ZhangXFetal2002}
X.-F. Zhang, D.~C. Cooper, F.~J. White, {Repeated cocaine treatment decreases
  whole-cell calcium current in rat nucleus accumbens neurons.}, J Pharmacol
  Exp Ther 301 (2002) 1119--25.

\bibitem{SchultzWetal97}
W.~Schultz, P.~Dayan, R.~Montague, {The computational role of dopamine D1
  receptors in working memory.}, Neural Networks 15 (2002) 561--72.

\bibitem{Schultz2002}
W.~Schultz, {Getting Formal with Dopamine and Reward.}, Neuron 36 (2002)
  241--263.

\bibitem{VittenIsaacson2001}
H.~Vitten, J.~S. Isaacson, {Synaptic transmission: exciting times for
  presynaptic receptors.}, Curr Biol 11 (2001) R695--7.

\bibitem{MarderEetal96}
E.~Marder, L.~F. Abbott, G.~G. Turrigiano, Z.~Liu, J.~Golowasch, {Memory from
  the dynamics of intrinsic membrane currents.}, Proc Natl Acad Sci U S A 93
  (1996) 13481--6.

\bibitem{GoldmanMSetal2001}
M.~S. Goldman, J.~Golowasch, E.~Marder, L.~F. Abbott, {Global structure,
  robustness, and modulation of neuronal models.}, J Neurosci 21 (2001)
  5229--38.

\bibitem{ChenXetal99}
X.~Chen, S.~B. Kombian, J.~A. Zidichouski, Q.~J. Pittman, {Dopamine depresses
  glutamatergic synaptic transmission in the rat parabrachial nucleus in
  vitro.}, Neuroscience 90 (1999) 457--68.

\bibitem{StenkampKetal98}
K.~Stenkamp, U.~Heinemann, D.~Schmitz, {Dopamine suppresses stimulus-induced
  field potentials in layer III of rat medial entorhinal cortex.}, Neurosci
  Lett 255 (1998) 119--21.

\bibitem{GaoWJetal2001}
W.~J. Gao, L.~S. Krimer, P.~S. Goldman-Rakic, {Presynaptic regulation of
  recurrent excitation by D1 receptors in prefrontal circuits.}, Proc Natl Acad
  Sci U S A 98 (2001) 295--300.

\bibitem{Law-ThoDetal94}
D.~Law-Tho, J.~C. Hirsch, F.~Crepel, {Dopamine modulation of synaptic
  transmission in rat prefrontal cortex: an in vitro electrophysiological
  study.}, Neurosci Res 21 (1994) 151--60.

\bibitem{HarveyLacey96}
J.~Harvey, M.~G. Lacey, {Endogenous and exogenous dopamine depress EPSCs in rat
  nucleus accumbens in vitro via D1 receptors activation.}, J Physiol 492
  (1996) 143--54.

\bibitem{UrbanNNetal2002}
N.~N. Urban, G.~Gonzalez-Burgos, D.~A. Henze, D.~A. Lewis, G.~Barrionuevo,
  {Selective reduction by dopamine of excitatory synaptic inputs to pyramidal
  neurons in primate prefrontal cortex.}, J Physiol 539 (2002) 707--12.

\bibitem{BehrJetal2000}
J.~Behr, T.~Gloveli, D.~Schmitz, U.~Heinemann, {Dopamine depresses excitatory
  synaptic transmission onto rat subicular neurons via presynaptic D1-like
  dopamine receptors.}, J Neurophysiol 84 (2000) 112--9.

\bibitem{HsuKSetal95}
K.~S. Hsu, C.~C. Huang, C.~H. Yang, P.~W. Gean, {Presynaptic D2 dopaminergic
  receptors mediate inhibition of excitatory synaptic transmission in rat
  neostriatum.}, Brain Res 690 (1995) 264--8.

\bibitem{HsuKS96}
K.~S. Hsu, {Characterization of dopamine receptors mediating inhibition of
  excitatory synaptic transmission in the rat hippocampal slice.}, J
  Neurophysiol 76 (1996) 1887--95.

\bibitem{EnnisMetal2001}
M.~Ennis, F.~M. Zhou, K.~J. Ciombor, V.~Aroniadou-Anderjaska, A.~Hayar,
  E.~Borrelli, L.~A. Zimmer, F.~Margolis, M.~T. Shipley, {Dopamine D2
  receptor-mediated presynaptic inhibition of olfactory nerve terminals.}, J
  Neurophysiol 86 (2001) 2986--97.

\bibitem{KogaMomiyama2000}
E.~Koga, T.~Momiyama, {Presynaptic dopamine D2-like receptors inhibit
  excitatory transmission onto rat ventral tegmental dopaminergic neurones.}, J
  Physiol 523 (2000) 163--73.

\bibitem{PisaniAetal2000}
A.~Pisani, P.~Bonsi, D.~Centonze, P.~Calabresi, G.~Bernardi, {Activation of
  D2-like dopamine receptors reduces synaptic inputs to striatal cholinergic
  interneurons.}, J Neurosci 20 (2000) RC69.

\bibitem{NicolaSMetal96}
S.~M. Nicola, S.~B. Kombian, R.~C. Malenka, {Psychostimulants depress
  excitatory synaptic transmission in the nucleus accumbens via presynaptic
  D1-like dopamine receptors.}, J Neurosci 16 (1996) 1591--604.

\bibitem{WangZetal2002}
Z.~Wang, X.~Q. Feng, P.~Zheng, {Activation of presynaptic D1 dopamine receptors
  by dopamine increases the frequency of spontaneous excitatory postsynaptic
  currents through protein kinase A and protein kinase C in pyramidal cells of
  rat prelimbic cortex.}, Neuroscience 112 (2002) 499--508.

\bibitem{MomiyamaSim96}
T.~Momiyama, J.~A. Sim, {Modulation of inhibitory transmission by dopamine in
  rat basal forebrain nuclei: activation of presynaptic D1-like dopaminergic
  receptors.}, J Neurosci 16 (1996) 7505--12.

\bibitem{FedericiMetal2002}
M.~Federici, S.~Natoli, G.~Bernardi, N.~B. Mercuri, {Dopamine selectively
  reduces GABA(B) transmission onto dopaminergic neurones by an unconventional
  presynaptic action.}, J Physiol 540 (2002) 119--28.

\bibitem{Gonzalez-IslasHablitz2001}
C.~Gonzalez-Islas, J.~J. Hablitz, {Dopamine inhibition of evoked IPSCs in rat
  prefrontal cortex.}, J Neurophysiol 86 (2001) 2911--8.

\bibitem{ShenJohnson2000}
K.~Z. Shen, S.~W. Johnson, {Presynaptic dopamine D2 and muscarine M3 receptors
  inhibit excitatory and inhibitory transmission to rat subthalamic neurones in
  vitro.}, J Physiol 525 (2000) 331--41.

\bibitem{DelgadoAetal2000}
A.~Delgado, A.~Sierra, E.~Querejeta, R.~F. Valdiosera, J.~Aceves, {Inhibitory
  control of the GABAergic transmission in the rat neostriatum by D2 dopamine
  receptors.}, Neuroscience 95 (2000) 1043--8.

\bibitem{NicolaMalenka97}
S.~M. Nicola, R.~C. Malenka, {Dopamine depresses excitatory and inhibitory
  synaptic transmission by distinct mechanisms in the nucleus accumbens.}, J
  Neurosci 17 (1997) 5697--710.

\bibitem{SeamansJKetal2001}
J.~K. Seamans, D.~Durstewitz, B.~R. Christie, C.~F. Stevens, T.~J. Sejnowski,
  {Dopamine D1/D5 receptor modulation of excitatory synaptic inputs to layer V
  prefrontal cortex neurons.}, Proc Natl Acad Sci U S A 98 (2001) 301--6.

\bibitem{CepedaCetal93}
C.~Cepeda, N.~A. Buchwald, M.~S. Levine, {Neuromodulatory actions of dopamine
  in the neostriatum are dependent upon the excitatory amino acid receptor
  subtypes activated.}, Proc Natl Acad Sci U S A 90 (1993) 9576--80.

\bibitem{SnyderGLetal2000}
G.~L. Snyder, P.~B. Allen, A.~A. Fienberg, C.~G. Valle, R.~L. Huganir, A.~C.
  Nairn, P.~Greengard, {Regulation of phosphorylation of the GluR1 AMPA
  receptor in the neostriatum by dopamine and psychostimulants in vivo.}, J
  Neurosci 20 (2000) 4480--8.

\bibitem{RayportSulzer95}
S.~Rayport, D.~Sulzer, {Visualization of antipsychotic drug binding to living
  mesolimbic neurons reveals D2 receptor, acidotropic, and lipophilic
  components.}, J Neurochem 65 (1995) 691--703.

\bibitem{FisherRSetal94}
R.~S. Fisher, M.~S. Levine, D.~R. Sibley, M.~A. Ariano, {D2 dopamine receptor
  protein location: Golgi impregnation-gold toned and ultrastructural analysis
  of the rat neostriatum.}, J Neurosci Res 38 (1994) 551--64.

\bibitem{ArianoSibley94}
M.~A. Ariano, D.~R. Sibley, {Dopamine receptor distribution in the rat CNS:
  elucidation using anti-peptide antisera directed against D1A and D3
  subtypes.}, Brain Res 649 (1994) 95--110.

\bibitem{ArianoMAetal92}
M.~A. Ariano, C.~J. Stromski, E.~M. Smyk-Randall, D.~R. Sibley, {D2 dopamine
  receptor localization on striatonigral neurons.}, Neurosci Lett 144 (1992)
  215--20.

\bibitem{vanderZeeEAetal95}
E.~A. Van~der Zee, J.~C. Compaan, B.~Bohus, P.~G. Luiten, {Alterations in the
  immunoreactivity for muscarinic acetylcholine receptors and colocalized PKC
  gamma in mouse hippocampus induced by spatial discrimination learning.},
  Hippocampus 5 (1995) 349--62.

\bibitem{vanderZeeEAetal97}
E.~A. van~der Zee, B.~Roozendaal, B.~Bohus, J.~M. Koolhaas, P.~G. Luiten,
  {Muscarinic acetylcholine receptor immunoreactivity in the amygdala--I.
  Cellular distribution correlated with fear-induced behavior.}, Neuroscience
  76 (1997) 63--73.

\bibitem{ScottLetal2002}
L.~Scott, M.~S. Kruse, H.~Forssberg, H.~Brismar, P.~Greengard, A.~Aperia,
  {Selective up-regulation of dopamine D1 receptors in dendritic spines by NMDA
  receptor activation.}, Proc Natl Acad Sci U S A 99 (2002) 1661--4.

\bibitem{LinRetal2001}
R.~Lin, K.~Karpa, N.~Kabbani, P.~Goldman-Rakic, R.~Levenson, {Dopamine D2 and
  D3 receptors are linked to the actin cytoskeleton via interaction with
  filamin A.}, Proc Natl Acad Sci U S A 98 (2001) 5258--63.

\bibitem{BindaAVetal2002}
A.~V. Binda, N.~Kabbani, R.~Lin, R.~Levenson, {D2 and D3 dopamine receptor cell
  surface localization mediated by interaction with protein 4.1N.}, Mol
  Pharmacol 62 (2002) 507--13.

\bibitem{ColginLLetal2003}
L.~L. Colgin, D.~Kubota, G.~Lynch, {Cholinergic plasticity in the
  hippocampus.}, Proc Natl Acad Sci U S A 100 (2003) 2872--7.

\bibitem{Koch99}
C.~Koch, {Biophysics of Computation:Information Processing in Single Neurons.}, Oxford University Press, 1999.

\bibitem{KonenWetal94}
W.~Konen, T.~Maurer, C.~von~der Malsburg, {A fast dynamic link mathing
  algorithm for invariant pattern recognition.}, Neural Networks 7 (1994)
  1019--1030.

\bibitem{Schmidhuber92}
J.~Schmidhuber, {Learning to control fast-weight memories: An alternative to
  dynamic recurrent networks.}, Neural Comput 4 (1992) 679--714.

\bibitem{SchelerSchumann2003}
G.~Scheler, J.~M. Schumann, Presynaptic modulation as fast synaptic switching:
  state-depende nt modulation of task performance, in: Proceedings of the 2003
  International Joint Conference on Neural Networks (IJCNN 2003), IEEE Press,
  2003.

\bibitem{Manahan-VaughanKulla2003}
D.~Manahan-Vaughan, A.~Kulla, {Regulation of Depotentiation and Long-term
  Potentiation in the Dentate Gyrus of Freely Moving Rats by Dopamine D2-like
  Receptors.}, Cereb Cortex 13 (2003) 123--35.

\bibitem{BlondOetal2002}
O.~Blond, F.~Crepel, S.~Otani, {Long-term potentiation in rat prefrontal slices
  facilitated by phased application of dopamine.}, Eur J Pharmacol 438 (2002)
  115--6.

\bibitem{AbrahamBear96}
W.~C. Abraham, M.~F. Bear, {Metaplasticity: the plasticity of synaptic
  plasticity.}, Trends Neurosci 19 (1996) 126--30.

\bibitem{JayTM2003}
T.~M. Jay, {Dopamine: a potential substrate for synaptic plasticity and memory
  mechanisms.}, Progress in Neurobiology 69 (2003) 375--390.

\bibitem{WickensKoetter94}
J.~Wickens, R.~Koetter, Cellular models of reinforcement, in: J.~C. Houk, J.~L.
  Davis, D.~G. Beiser (Eds.), Models of Information Processing in the Basal
  Ganglia, MIT, 1994, pp. 187--214.

\bibitem{Magee2003}
J.~C. Magee, {A prominent role for intrinsic neuronal properties in temporal
  coding.}, Trends Neurosci 26 (2003) 14--6.

\bibitem{SalinasSejnowski2001}
E.~Salinas, T.~J. Sejnowski, {Gain modulation in the central nervous system:
  where behavior, neurophysiology, and computation meet.}, Neuroscientist 7
  (2001) 430--40.

\bibitem{EgorovAVetal2002}
A.~V. Egorov, B.~N. Hamam, E.~Fransen, M.~E. Hasselmo, A.~A. Alonso, {Graded
  persistent activity in entorhinal cortex neurons.}, Nature 420 (2002) 173--8.

\bibitem{LiuYetal99}
Y.~Liu, D.~E. Krantz, C.~Waites, R.~H. Edwards, {Membrane trafficking of
  neurotransmitter transporters in the regulation of synaptic transmission.},
  Trends Cell Biol 9 (1999) 356--63.

\bibitem{BlakelyBauman2000}
R.~D. Blakely, A.~L. Bauman, {Biogenic amine transporters: regulation in
  flux.}, Curr Opin Neurobiol 10 (2000) 328--36.

\bibitem{LetchworthSRetal2001}
S.~R. Letchworth, M.~A. Nader, H.~R. Smith, D.~P. Friedman, L.~J. Porrino,
  {Progression of changes in dopamine transporter binding site density as a
  result of cocaine self-administration in rhesus monkeys.}, J Neurosci 21
  (2001) 2799--807.

\bibitem{FransenEetal2002}
E.~Fransen, A.~A. Alonso, M.~E. Hasselmo, {Simulations of the role of the
  muscarinic-activated calcium-sensitive nonspecific cation current INCM in
  entorhinal neuronal activity during delayed matching tasks.}, J Neurosci 22
  (2002) 1081--1097.

\end{thebibliography}

\end{document}